\documentclass[journal]{IEEEtran}
\newif\ifsubmitversion
\newif\ifarxivversion
\arxivversiontrue
\submitversiontrue

\usepackage{graphicx}
\graphicspath{{./fig/epspdf/}{./fig/jpg/}}
\DeclareGraphicsExtensions{.pdf,.jpeg,.png}

\usepackage{amsmath,amsfonts,amsthm,amssymb}

\usepackage{accents}
\newlength{\dhatheight}

\usepackage{multirow}
\usepackage{array}
\usepackage{cite}

\usepackage{color}
\definecolor{lightblue}       {rgb}{0.00,0.10,1.00}
\definecolor{royalblue}       {rgb}{0.23,0.40,1.00}
\definecolor{darkbluem}       {rgb}{0.06,0.31,0.66} 
\definecolor{blue(pigment)}   {rgb}{0.20,0.20,0.60}
\definecolor{mediumaquamarine}{rgb}{0.40,0.80,0.67}
\definecolor{tealgreen}       {rgb}{0.00,0.51,0.50}
\definecolor{denim}           {rgb}{0.08,0.38,0.74}
\definecolor{cornflower}      {rgb}{0.40,0.55,0.90}
\definecolor{darkblue}        {rgb}{0.27,0.23,0.54}
\definecolor{darkbrown}       {rgb}{0.65,0.16,0.16}
\definecolor{darkgreen}       {rgb}{0.13,0.54,0.13}
\definecolor{darkorange}      {rgb}{1.00,0.50,0.00}
\definecolor{carminepink}     {rgb}{0.92,0.30,0.26}
\definecolor{darkpink}        {rgb}{1.00,0.08,0.57}
\definecolor{darkred}         {rgb}{0.68,0.13,0.13}

\definecolor{byzantium}{rgb}{0.44, 0.16, 0.39}

\def\firstbest#1{\begin{color}{black}\textbf{#1}\end{color}}%
\def\secondbest#1{\begin{color}{blue(pigment)}\textbf{#1}\end{color}}%
\def\thirdbest#1{\begin{color}{byzantium}\textbf{#1}\end{color}}%

\def\firstbest#1{\begin{color}{black}\textbf{#1}\end{color}}%
\def\secondbest#1{\begin{color}{denim}{#1}\end{color}}%
\def\thirdbest#1{\begin{color}{darkpink}{#1}\end{color}}%

\usepackage{mathtools}
\DeclarePairedDelimiterX{\norm}[1]{\lVert}{\rVert}{#1}

\ifCLASSOPTIONcompsoc
  \usepackage[caption=false,font=normalsize,labelfont=sf,textfont=sf]{subfig}
\else
  \usepackage[caption=false,font=footnotesize]{subfig}
\fi

\usepackage{stfloats}
\usepackage{units}

\usepackage{marginnote}

\usepackage{textcomp}

\usepackage{algpseudocode}
\usepackage{algorithm}

\def\mm#1{\boldsymbol{\mr{#1}}}
\def\mmh#1{\mm{\hat{#1}}}
\def\mmhh#1{\mm{\doublehat{#1}}}

\def\mmhh#1{\widehat{\widehat{\mm{#1}}}}
\def\mmhh#1{\mm{\breve{#1}}}

\def\mv#1{\boldsymbol{{#1}}}
\def\mvh#1{\mv{\hat{#1}}}
\def\textem#1{{\em{#1}}}%
\def\mb#1{\mbox{\boldmath $#1$}}
\def\mbb#1{{\mathbb{#1}}}
\def\mc#1{{\mathcal{#1}}}
\def\mr#1{{\mathrm{#1}}}

\def\lrb#1{\ensuremath{\left({#1}\right)}}%
\def\lrc#1{\ensuremath{\left\{{#1}\right\}}}%
\def\lrs#1{\ensuremath{\left[{#1}\right]}}%

\def\tran#1{{#1}^{\kern-1.5pt{\mathsf{T}}}}%

\def\br#1{\boldsymbol{\mr{#1}}}

\def\mxm#1#2{\ensuremath{{{#1}}\hspace{0.2pt}{\times}\hspace{0.2pt}{{#2}}}}%
\def\sxs#1#2{\ensuremath{{\text{#1}}\hspace{-1pt}{\times}\hspace{-1pt}{\text{#2}}}}%

\def\qq#1{``#1''}%

\makeatletter
\newcommand{\replunderscores}[1]{\expandafter\@repl@underscores#1_\relax}%

\def\@repl@underscores#1_#2\relax{%
    \ifx \relax #2\relax
        #1%
    \else
        #1%
        \textunderscore
        \@repl@underscores#2\relax
    \fi
}%
\makeatother

\def\textus#1{\replunderscores{#1}}%

\def\hadamard#1#2{\ensuremath{{#1}\raisebox{0.3ex}{\scriptsize$\odot$}{#2}}}%

\DeclareMathSymbol{\shortminus}{\mathbin}{AMSa}{"39}%

\def\MS#1{\ensuremath{\mm{M}^{(#1)}}}

\def\REMOVED#1{}

\begin{document}
\title{Speech Enhancement Using Multi-Stage Self-Attentive Temporal Convolutional Networks}
\author{Ju Lin, \IEEEmembership{Student Member, IEEE, } 
        Adriaan J.\ van Wijngaarden, \IEEEmembership{Senior Member, IEEE, } 
        \hbox{Kuang-Ching Wang, \IEEEmembership{Senior Member, IEEE, }}  
        and 
        Melissa C. Smith, \IEEEmembership{Senior Member, IEEE} 
\thanks{The site \textus{https://linjucs.github.io/demo/ms\_sa\_tcn.html} contains 
audio samples that demonstrate the effectiveness of the proposed multi-stage SA-TCN
speech enhancement systems.}
\thanks{This work is supported by the US Army Medical Research and Materiel Command under 
Contract No.\ W81XWH-17-C-0238.}
\thanks{Ju Lin, Kuang-Ching Wang and Melissa C. Smith are with the Department of Electrical 
        and Computer Engineering, Clemson University, Clemson, SC 29634, USA 
       (e-mail: \{jul, kwang, smithmc\}@clemson.edu).}
\thanks{Adriaan J.\ de Lind van Wijngaarden is with Nokia Bell Labs, Murray Hill, NJ 07974, 
        USA (e-mail: adriaan.de\_lind\_van\_wijngaarden@nokia-bell-labs.com).}}

\ifarxivversion
\else
\markboth{IEEE/ACM Trans. Audio, Speech, and Language Proc. - initial submission,
Feb.\ 23, 2021}%
{Ju \MakeLowercase{\textit{et al.}}: Speech Enhancement Using Multi-Stage Self-Attentive Temporal Convolutional Networks}
\IEEEpubid{0000--0000/00\$00.00~\copyright~2021 IEEE}
\fi
\IEEEaftertitletext{\vspace{-2\baselineskip}}
\maketitle

\begin{abstract}
  Multi-stage learning is an effective technique to invoke multiple
  deep-learning modules sequentially. This paper applies multi-stage
  learning to speech enhancement by using a multi-stage structure, where
  each stage comprises a self-attention (SA) block followed by stacks of
  temporal convolutional network (TCN) blocks with doubling dilation
  factors. Each stage generates a prediction that is refined in a subsequent
  stage. A fusion block is inserted at the input of later stages to
  re-inject original information. The resulting multi-stage speech
  enhancement system, in short, multi-stage SA-TCN, is compared with
  state-of-the-art deep-learning speech enhancement methods using the
  LibriSpeech and VCTK data sets. The multi-stage SA-TCN system's
  hyper-parameters are fine-tuned, and the impact of the SA block, the fusion
  block and the number of stages are determined. The use of a multi-stage
  SA-TCN system as a front-end for automatic speech recognition systems is
  investigated as well. It is shown that the multi-stage SA-TCN systems 
  perform well relative to other state-of-the-art systems in terms of speech 
  enhancement and speech recognition scores.
\end{abstract}


\section{Introduction}
\IEEEPARstart{S}{peech} enhancement is a basic function that is used to improve 
the quality and the intelligibility of a speech signal that is degraded by
ambient noise. Speech enhancement algorithms are used extensively in many
audio- and communication systems, including mobile handsets, speaker
verification systems and hearing aids. Popular classic techniques include
spectral-subtraction algorithms, statistical model-based methods that use
maximum-likelihood (ML) estimators, Bayesian estimators, minimum mean
squared error (MMSE) methods, subspace algorithms based on single value
decomposition and noise-estimation algorithms (see~\cite{loizou2013speech}
and references therein). Modern techniques often use deep learning.
Early examples include a recurrent neural network (RNN) to model long-term acoustic
characteristics~\cite{maas2012recurrent}, and a deep auto-encoder that denoises
speech signals with greedy layer-oriented pre-training~\cite{lu2013speech}. 
In~\cite{xu2015regression}, a deep neural network (DNN) was used as a non-linear regression function.
In~\cite{tan2018convolutional}, a convolutional recurrent neural network
(CRN) was used, consisting of a convolutional encoder-decoder architecture
and multiple long short-term memory (LSTM) layers that aim 
to capture long-context information. Other speech enhancement systems use
a generative adversarial network (GAN), which is known for its ability to 
generate natural-looking signals in the time or frequency domain~\cite{pascual2017segan, 
fu2019metricgan, LNWxx19,baby2019sergan, lin2020improved}. Recent
studies consider the use of an attention mechanism~\cite{giri2019attention, kim2020t, 
zhao2020monaural, koizumi2020speech, pandey2020dense, zhao2020noisy}. 
Self-attention~\cite{vaswani2017attention} is an efficient context information
aggregation mechanism that operates on the input sequence itself and that
can be utilized for any task that has a sequential input and 
output. In~\cite{pandey2020dense}, self-attention is combined with a dense
convolutional neural network. A time-frequency (T-F) attention method,
proposed in~\cite{zhao2020noisy}, combines time-domain and frequency-domain
attention to perform denoising and dereverberation at the same time.

A temporal convolutional network (TCN) consists of dilated 1-D
convolutions that create a large temporal receptive field with fewer parameters
than other models. Recent research shows that TCN-based models achieve excellent 
performance for text-to-speech~\cite{ODZxx16_D}, speech enhancement~\cite{pandey2019tcnn,
zhang2019monaural, koyama2020exploring, KTP20, liu2020multichannel}, and speech
separation~\cite{luo2019conv}. In~\cite{zhang2019monaural}, a speech
enhancement system was proposed that uses a multi-branch TCN, in short
MB-TCN, which effectively performs a split-transform-aggregate operation and
enables the model to learn and determine an accurate representation by
aggregating the information from each branch. In~\cite{KTP20}, the TCN used
in~\cite{luo2019conv} for speech separation was adapted for speech enhancement 
and integrated in a multi-layer encoder-decoder architecture. The use of a 
complex Short-Time Fourier transform (STFT) for TCN-based speech enhancement 
rather than magnitude or time-domain features was investigated
in~\cite{koyama2020exploring}.
\IEEEpubidadjcol

The above-mentioned methods can generally be classified as feature-mapping and 
mask-learning methods, which are two commonly used deep-learning approaches for 
single-channel speech enhancement methods for stereo data. Feature mapping 
approaches enhance the noisy features using a mapping network that minimizes the 
mean square error between the enhanced and clean features. Mask-learning approaches 
estimate the ideal ratio mask, the ideal binary mask or the complex ratio mask, and 
then use this mask to 
filter noisy speech signals and reconstruct the clean speech signals. Mask-learning 
methods usually perform better than feature mapping methods in terms of speech 
quality metrics~\cite{wang2014training,chen2017improving, narayanan2014investigation}.

Recently, multi-stage learning has been successfully applied for a wide
variety of tasks, including human pose estimation~\cite{newell2016stacked}, 
action segmentation~\cite{farha2019ms}, speech
enhancement~\cite{hao2020masking, li2020two, li2020recursive} 
and speech separation~\cite{fan2020deep}. A multi-stage architecture 
consists of stages that sequentially use the same model or a combination
of different models, and each model operates directly on the output of the previous
stage. The effect of such an arrangement is that the model used in a given
stage takes the predictions from prior stages as input and incrementally
refines these predictions. 

Multi-stage learning systems that perform the same task in each stage
typically use the same supervision principles in each intermediate
stage~\cite{newell2016stacked, farha2019ms, li2020recursive}.
In~\cite{farha2019ms}, multiple stacked TCN networks are proposed for action
segmentation. In~\cite{li2020recursive}, a multi-stage network with dynamic
attention is introduced, where the intermediate output in each stage is
corrected with a memory mechanism. To reduce the model parameters, each
stage uses a shared network. It is shown that this multi-stage approach
typically performs better than systems with a larger and deeper network.

Multi-stage learning systems where each stage performs a different task are
considered in~\cite{hao2020masking, li2020two, fan2020deep}. Here, each
stage has a different task and a different target. The performance can be 
improved by aggregating different stages if the nature of each stage is 
complementary. For instance, a two-stage speech enhancement approach is 
presented in~\cite{hao2020masking}, where the first stage uses a model 
to predict a binary mask to remove frequency bins that are dominated 
by severe noise, and where the second stage performs in-painting of the 
masked spectrogram from the first stage to recover the
speech spectrogram that was removed in the first stage. In~\cite{li2020two},
a two-stage algorithm is proposed to optimize the magnitude and phase
separately. The magnitude is optimized in the first stage and the enhanced
magnitude and phase are then further refined jointly.

This paper details a novel multi-stage speech enhancement system, where each
stage comprises a self-attention (SA) block~\cite{vaswani2017attention}
followed by stacks of dilated temporal convolutional network (TCN) blocks.
The system is referred to as a multi-stage SA-TCN speech
enhancement system. Each stage generates a prediction in the form of a soft
mask that is refined in each subsequent stage. Each self-attention block
produces a dynamic representation for different noise environments and their
relevance across frequency bins, as such enhancing the features, and the
stacks of TCN blocks perform sequential refinement processing. A fusion
block is inserted at the input of later stages to re-inject original speech
information to mitigate possible speech information loss in earlier stages.

This paper is organized as follows. Section~\ref{sec:ms-tcn} details the
proposed multi-stage SA-TCN speech enhancement system and the underlying SA,
TCN, and fusion blocks. Section~\ref{sec:experiment} details the
comprehensive experiments using the LibriSpeech~\cite{panayotov2015librispeech} and
VCTK~\cite{valentini2016investigating} corpus. Section~\ref{sec:results}
first presents the experiments that were performed to fine-tune the multi-stage 
SA-TCN system's hyper-parameters, to determine the optimum number of stages, and 
to quantify the impact of the SA block and the fusion block on the performance. 
The use of the proposed multi-stage SA-TCN system as a front-end for automatic speech
recognition (ASR) systems is investigated as well. Extensive experiments 
with the LibriSpeech~\cite{panayotov2015librispeech} and VCTK~\cite{valentini2016investigating} 
corpus show that multi-stage SA-TCN systems achieve significantly better speech
enhancement and speech recognition scores than other state-of-the-art speech
enhancement systems. Section~\ref{sec:concl} concludes the paper and
discusses further research directions.

\section{Multi-Stage SA-TCN Systems}%
\label{sec:ms-tcn}%
Speech enhancement systems take a sampled received noisy speech signal as
input and aim to reconstruct the speech signal. Let $\lrc{x(t) | t \in
\mbb{Z}}$ denote a deterministic discrete-time data sequence that is
obtained by sampling a received continuous-time noisy speech signal
$x_c(\cdot)$ at time interval $T_s$, i.e., $x(t) = x_c(t \cdot T_s) \in
\mbb{R}$, and let the total number of samples be denoted by $T_{\ell}$. The
short-time Fourier transform (STFT) of length $N$ of $\lrc{x(t)}$ with
window function $w(t)$ of length $N$ and hop-length $T_h$ is given by
\begin{equation}
  X_{\tau,\omega} = \sum_{n=0}^{N-1} w(n) x(\tau{T_h} + n) \exp\lrb{-j\frac{2\pi\omega{n}}{N}},
\end{equation}
where $\tau$ is the index of the sliding window and $0 \le \omega < N$ is
the frequency index. In this paper, a Hanning window is used, where 
\begin{equation}
  w(t) = \sin^2\lrb{\frac{\pi{t}}{N-1}}.
\end{equation}
Let $\mm{X}$ and $\mm{\Omega}$ denote the STFT magnitude and phase, i.e., 
$\mm{X} = \lrc{|X_{\tau,\omega}|} \in \mbb{R}^{F{\times}T}$ and $\mm{\Omega} \in \mbb{R}^{F{\times}T}$, 
where $F = N/2 + 1$ denotes the number of frequency bins and $T = T_{\ell}/{T_h} + 1$. 

\begin{figure*}[!ht]
  \centering
  \centerline{\includegraphics[width=0.99\textwidth]{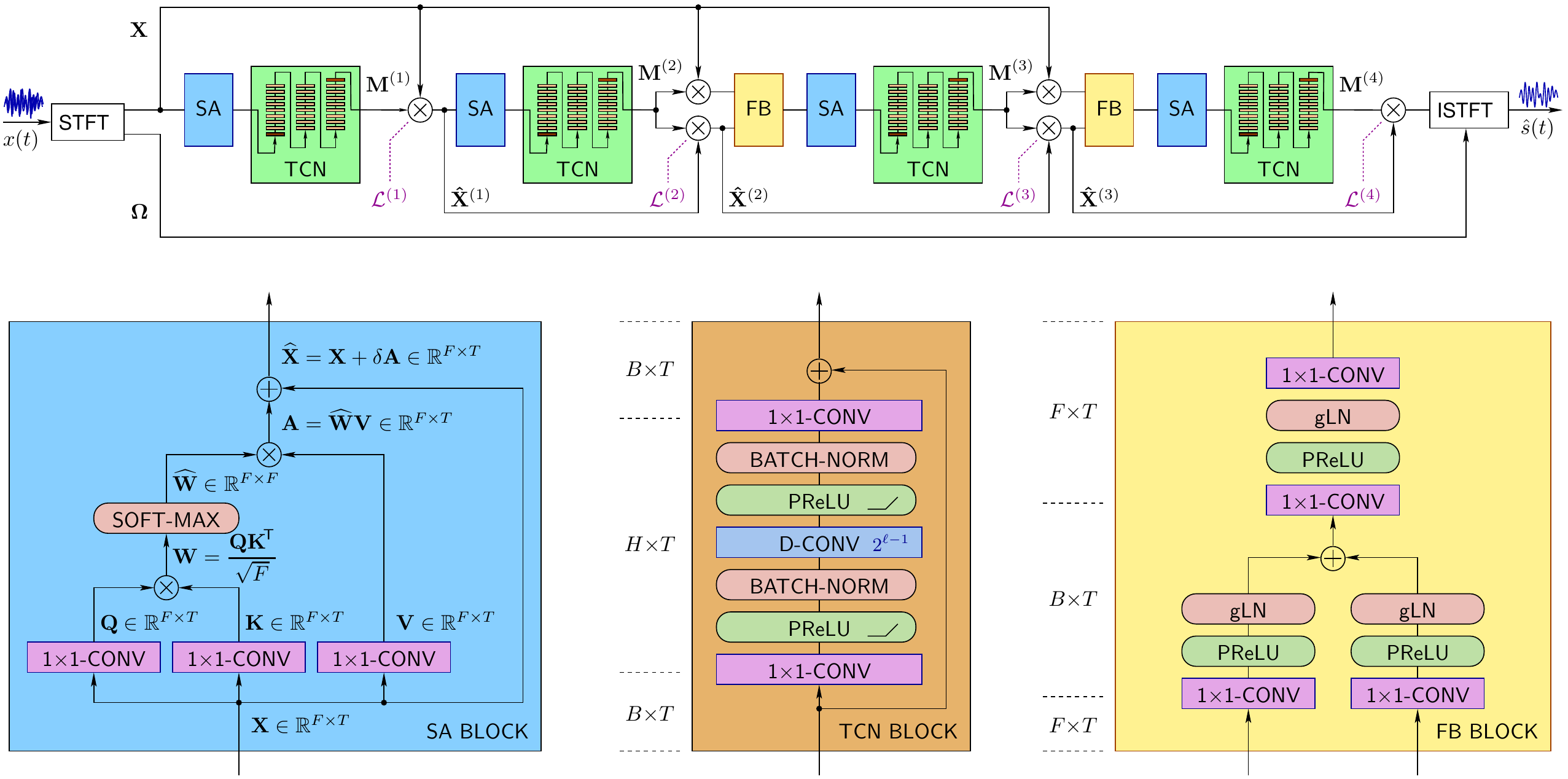}}
  \caption{Block diagram of a multi-stage SA-TCN speech enhancement system
    with $K=4$ stages, where each stage consists of a self-attention (SA) block, 
    followed by $R=3$ stacks of $L=8$ TCN blocks, where the dilation factor
    of the $\ell$-th TCN block in the stack equals $\Delta_{\ell} = 2^{\ell-1}$, i.e., the
    dilation doubles for each next TCN block in the stack. A fusion block is used as 
    of Stage~3 to re-inject the original STFT magnitude. The figure also shows the 
    detailed structure  
    of the self-attention module, with frequency and time dimensions $F$ and
    $T$, a single TCN block with hyper-parameters $B$ and $H$ and $P$, and
    the proposed fusion block.}
\label{fig:ms-tcn}
\end{figure*}

The proposed multi-stage SA-TCN speech enhancement system consists of $K$ stages. Fig.~\ref{fig:ms-tcn}
illustrates a 4-stage SA-TCN system. Each stage comprises a self-attention (SA) block
followed by $R$ stacks of $L$ TCN blocks. For $K$-stage SA-TCN systems where $K \ge 3$, a feature 
fusion block is inserted prior to each stage $k$, where $3 \le k \le K$.

Each of the blocks have special features that are particularly suited for
speech enhancement. The self-attention mechanism aggregates context
information across channels, which is particularly helpful in obtaining a
dynamic representation when the noise is non-stationary, and this is the case 
for many speech enhancement scenarios.

The TCN consists of $R$ stacks of $L$ non-causal TCN blocks, where the dilation 
factor of the $\ell$-th TCN block in the stack is given by $\Delta_{\ell} = 2^{\ell}$. 
As such, each stack has a large receptive field, which makes it particularly suited for 
temporal sequence modeling. Each TCN block has a skip connection between the input and 
output to reduce the loss of low-level details and to provide hooks for optimization.

The multi-stage architecture iteratively refines the initial predictions. It 
should be noted that the prediction of a previous stage may include some errors. For 
instance, the frequency bins dominated by speech may be masked and the resulting magnitude 
spectrogram may have lost some of the speech information. A fusion block block is 
inserted prior to each stage $k$, where $3 \le k \le K$, that combines the predicted 
magnitude $\mmh{X}{}^{(k-1)}$ at the output of stage $k-1$ and the original magnitude 
$\mm{X}$ as input, in order to re-inject the original speech information.
 
The first stage consists of a self-attention (SA) block that takes $\mm{X}$
as input and that uses three {\sxs{1}{1}}-convolutions to form the query $\mm{Q}$ and 
the key-value pair $(\mm{K},\mm{V})$, where $\mm{Q}, \mm{K}, \mm{V} \in \mbb{R}^{F \times
T}$. In order to compute the attention component $\mm{A}$, we first compute the weight
$\mm{W}$, given by 
\begin{equation}
  \mm{W} = \frac{\mm{Q}\tran{\mm{K}}}{\sqrt{F}},
\end{equation}
and then use the soft-max function $\sigma(\cdot)$ to obtain
$\mm{\widehat{W}} = \{\widehat{W}_{i,j}\} = \sigma\lrb{\mm{W}}$, i.e., 
\begin{equation}
  \widehat{W}_{i,j} = \frac{\exp\lrb{W_{i,j}}}{w_j}, \text{\ where \ } w_j = \sum_{i=1}^{F}\exp\lrb{W_{i,j}}.
\end{equation}

The attention component $\mm{A} \in \mbb{R}^{F \times T}$ is now determined using 
\begin{equation}
  \mm{A} = \mm{\widehat{W}}\mm{V}, 
\end{equation}
The SA block outputs $\mm{\widehat{X}} = \mm{X} + \delta\mm{A}$, where $\delta$
is a scalar with initial value zero that is used to allow the network to first rely on
the cues in the local channels $\mm{X}$ and then gradually assign more weight to the 
non-local channels using back-propagation to reach its optimal value.

The output $\mm{\widehat{X}} \in \mbb{R}^{F \times T}$ is fed into a TCN with input 
feature dimension $B$ and network feature map dimension $H$ by using a bottleneck layer 
to reduce the number of channels from $F$ to $B$. The TCN consists of $R$ identical stacks 
of $L$ TCN blocks. Each TCN block comprises an \sxs{1}{1} convolution at its input to match the 
input feature dimension $B$ to the TCN block's internal feature map dimension $H$, a dilated 
depth-wise convolution (D-conv) layer with kernel size $P$ and dilation factor
$\Delta_{\ell} = 2^{\ell-1}$, 
where $\ell$ denotes the order of the TCN block in the stack of $L$ TCN blocks, and a 
\sxs{1}{1} convolution layer to reduce the number of channels at the output from $H$ to $B$.
This output is then recombined with the input using a skip connection to avoid losing 
low-level details. A parametric rectified linear unit (\textproc{PReLU}) activation 
layer~\cite{he2015delving} and a batch normalization layer~\cite{ioffe2015batch} are inserted
prior to and after the depth-wise convolution layer to accelerate training and improve 
performance. A sigmoid function is applied at the output of the last TCN block of the
last stack to obtain a [0-1] mask $\MS{1}$ that minimizes the mean absolute
error loss 
\begin{equation}
  \mc{L}^{(1)} = \norm{\hadamard{\MS{1}}{\mm{X}}-\mm{S}},
\end{equation} 
where the operator $\hadamard{}{}$ denotes the Hadamard product and $\mm{S}$
denotes the STFT magnitude of the clean speech signal $s(t)$. 

The stack of $L$ TCN blocks with kernel $P$ and dilation factor
$\Delta_{\ell} = 2^{\ell-1}$  create a receptive field of size $R^{(P,L)}$, given by
\begin{equation}
  R^{(P,L)} = 1 + \sum_{\ell = 1}^{L} (P-1) \cdot 2^{\ell-1}. 
\end{equation}
As such, a stack of $L$ TCN blocks creates a large temporal receptive field with 
fewer parameters than other models.

\begin{figure}[!ht]
  \begin{center}
    \begin{tabular}{cc}
      \includegraphics[width=0.97\columnwidth]{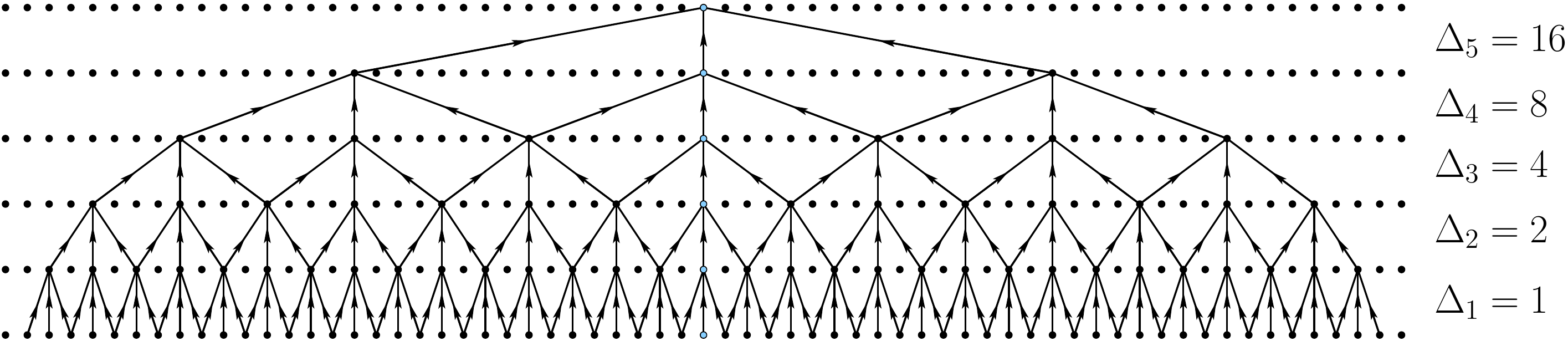}
    \end{tabular}
  \end{center}
  \vspace{-2mm}
  \caption{\label{fig:mos}  
    Example of the connections formed by a non-causal D-convolution for a stack of $L = 5$ TCN blocks, each 
    with kernel $P=3$ and dilation $\Delta_{\ell} = 2^{\ell-1}$, where $1 \le \ell \le L$.
  }
  \label{fig:dilation}
\end{figure}

This paper considers multi-stage SA-TCN systems with kernel size $P=3$. An illustration of the 
receptive field for a stack of $L=5$ TCN blocks with kernel size $P=3$ is shown in 
Fig.~\ref{fig:dilation}. 
The multi-stage SA-TCN system's hyper-parameters $(B, H, R, L)$ will be
optimized using experiments.

As indicated, the same SA-TCN structure is used for subsequent stages, and an
additional element, a fusion block, is inserted prior to each stage if there
are three or more stages.

For notational convenience, let $\br{\Psi}_{k}^{(R,L)}(\cdot)$ denote the
mapping performed by the $R$ stacks of $L$ TCN blocks in stage $k$, and let
$\br{\Upsilon}_{k}(\cdot)$ denote the self-attention operation at stage
$k$. It follows that $\MS{1}$ can now be expressed as
\begin{equation}
  \MS{1} = S(\mm{\Psi}_{1}^{(R,L)}(\mm{\Upsilon}_{1}(\mm{X}))),
\end{equation}
where $S(\cdot)$ denotes the sigmoid function. As such, $\MS{1}$ is the predicted mask 
at the output of the first stage. The enhanced speech STFT magnitude
$\mmh{X}{}^{(1)}$ at the output of stage 1 is given by $\mmh{X}{}^{(1)} =
\hadamard{\MS{1}}{\mm{X}}$.

In a similar fashion, the predicted mask $\MS{2}$ at the output of the second stage
can be obtained by evaluating
\begin{equation}
  \MS{2} = S(\mm{\Psi}_{2}^{(R,L)}(\mm{\Upsilon}_{2}({\mmh{X}{}^{(1)}}))) 
\end{equation}
and the estimated STFT magnitude $\mmh{X}{}^{(2)} =
\hadamard{\MS{2}}{\mmh{X}{}^{(1)}}$.

A multi-stage SA-TCN speech enhancement system with three or more stages ($K
\ge 3$) is constructed by inserting
a fusion block that performs operation $\mm{\Phi}(\cdot)$ prior to each stage $k$, 
where $3 \le k \le K$, taking the masked STFT magnitude $\mmh{X}{}^{(k-1)}$ and 
STFT magnitude $\mm{X}$ as inputs. Each input is passed through a {\sxs{1}{1}}-convolution 
and a \textproc{PReLU} operation, after which a global layer normalization
(\textproc{gLN}) is performed~\cite{luo2019conv}. The operation $\textproc{gLN}(\mm{Y})$ 
is given by   
\begin{equation}
  \textproc{gLN}(\mm{Y})   
     = \frac{\mm{Y} - E\lrs{\mm{Y}}}
             {\sqrt{\text{var}(\mm{Y}) 
               + \epsilon}}\odot\mb{\gamma} + \mb{\beta}, 
\end{equation}
where $\mm{Y}\in\mbb{R}^{{F}\times{T}}$ is the input feature with mean
$E\lrs{\mm{Y}}$ and variance $\text{var}(\mm{Y})$, $\mb{\gamma},
\mb{\beta} \in\mbb{R}^{{F}\times{1}}$ are trainable parameters, 
and $\epsilon$ is a small constant for numerical
stability. 

The outputs of the two \textproc{gLN} are added, and the result is again
sent through a {\sxs{1}{1}}-convolution, a \textproc{PReLU}, another
\textproc{gLN}, another {\sxs{1}{1}}-convolution and another
\textproc{PReLU}. The output, denoted as $\mmhh{X}{}^{(k-1)}$, is given by
\begin{equation}
  \mmhh{X}{}_{k-1} = \mm{\Phi}_{k}\left(
                     \hadamard{\MS{k-1}}{\mm{X}},{\mmh{X}{}^{(k-1)}}%
                     \right).%
\end{equation}
The output $\mmhh{X}{}_{k-1}$ is then used as an input to the next stage, and the 
expression for the mask $\MS{k}$ at the output of the $k$-th stage is now
given by 
\begin{equation}
  \MS{k} = S\bigl(%
              \mm{\Psi_k}^{(R,L)}\bigl(%
                \mm{\Upsilon_k}\bigl(%
                \mmhh{X}{}_k
               \bigr)%
               \bigr)%
              \bigr).
\end{equation}

The enhanced magnitude $\mmh{X}{}^{(k)}$  at stage $k$ is given by
\begin{equation}
  \mmh{X}^{(k)} = \hadamard{\MS{k}}{\mmh{X}{}^{(k-1)}}.%
\end{equation}

Each next stage $k$, where $k > 1$, computes mask $\MS{k}$ that minimizes
the mask-based signal approximation mean absolute error loss $\mc{L}^{(k)}$ using
\begin{equation} \mc{L}^{(k)} =
  \norm[\big]{\hadamard{\MS{k}}{\mmh{X}{}^{(k-1)}}-\mm{S}},
\end{equation}
where $\mmh{X}{}^{(k-1)}$ denotes the estimated STFT magnitude at stage $k-1$.

At the output of the last stage of the multi-stage SA-TCN system, the time-domain waveform 
$\mvh{s}$ is computed using the processed STFT magnitude $\mmh{X}{}^{(k)}$ and the 
original STFT phase $\Omega$ by applying the inverse
STFT, in short ISTFT, denoted as
\begin{equation}
  \mvh{s} = \text{ISTFT}(\mmh{X}{}^{(K)}, \mm{\Omega}).
\end{equation}

The proposed multi-stage SA-TCN system provides a mean absolute error loss
$\mc{L}^{(k)}$ at the output of each stage. Since each stage provides an equal contribution 
during the training process, we use the accumulated mask-based signal approximation training 
objective function
\begin{equation}
  \mc{L} = \sum_{k=1}^{K} \mc{L}^{(k)}.
\end{equation}
The use of the mean absolute error loss is motivated by recent observations
that it achieves better objective quality scores when using spectral mapping
techniques~\cite{pandey2018adversarial,pandey2019new}. 

\section{Experimental Setup}%
\label{sec:experiment}
In the following, the data set, model set up and the evaluation metrics are
detailed.

\subsection{Data Set}
To verify the effectiveness of the proposed multi-stage SA-TCN system, we conduct 
experiments using the LibriSpeech and VCTK data sets. The
detailed set-up for each data set is detailed below.

\textbf{LibriSpeech} is an open-source corpus that contains 960 hours of speech
derived from audio books in the LibriVox project. The sampling frequency is
\unit[16]{kHz}. The clean source is trained using 100 hours of speech data from 
the \qq{train-clean} data set. The validation set uses 800 sentences from 
the \qq{dev-clean} data set, and the test set uses 500 sentences from 
the \qq{test-clean} data set. The training set uses 10,000 randomly selected noise 
sample sequences from the DNS Challenge~\cite{reddy2020icassp}. 
The training clean speech has been cut to 75,206 4-second segments. The 
training and validation sets distort the clean segments with a randomly-selected 
noise sound from the DNS Challenge noise set with an SNR in the set $\{-5,
-4, \cdots, 9, 10\}$ (in dB), The test set uses three distinct noise types: 
\qq{babble noise} from the NOISEX-92 corpus~\cite{varga1993assessment}, and 
\qq{office noise} and \qq{kitchen noise} from the DEMAND noise 
corpus~\cite{thiemann2013diverse}. The first channel signal of the corpus is 
used for data generation. Each clean utterance is distorted by a randomly 
selected noise type at a randomly selected SNR from the set $\{-5, 0, 5, 10, 15\}$ 
(in dB).

The \textbf{VCTK} database used here is derived from the Valentini-Botinhao
corpus~\cite{valentini2016investigating}. Each speaker fragment contains
about 10 different sentences. The training set uses 28 speakers, and the
test set uses two speakers. The training set used here uses 40 noise
conditions: eight noise types and two artificial noise types from the Demand
database~\cite{thiemann2013diverse}) are used at a randomly selected SNR
from the set $\{0, 5, 10, 5\}$ (in dB). The test set uses 20 noise
conditions: five noise types from the Demand database at a randomly selected
SNR from the set $\{2.5, 7.5, 12.5, 17.5\}$ (in dB). There are about 20
different sentences in each condition for each test speaker. The test set
conditions are different from the training set, as the test set uses
different speakers and noise conditions.

\subsection{Model Setup}
The baseline systems used for performance comparison are a CRN system~\cite{tan2018convolutional}, a 
complex-CNN system that is based on concepts proposed
in~\cite{wang2018voicefilter} and that was adapted for speech enhancement, and a multi-stage system 
DARCN~\cite{li2020recursive}. The setup of the baseline systems and the proposed multi-stage SA-TCN 
systems are detailed below.

\noindent \textbf{CRN:} The CRN-based approach takes the magnitude as input. Instead of
directly mapping the noisy magnitude to the clean magnitude, we adapted the CRN to predict 
the ratio mask and as such improve its performance. The CRN-based method consists of five 2D convolution layers with filters of 
size \mxm{3}{2} each and [16, 32, 64, 128, 256] output channels, respectively.  
This output is post-processed by two LSTM layers with 1024 nodes each, and five
2D deconvolution layers with filter size \mxm{3}{2} each and output 
channels [128, 64, 32, 16, 1], respectively. 

\noindent \textbf{Complex-CNN.} 
The complex-CNN performs a complex spectral mapping~\cite{fu2017complex, tan2019complex}, where
the real and imaginary spectrograms of the noisy speech signal are treated as two different input 
channels. An STFT is used with a \unit[20]{ms} Hanning window, a \unit[20]{ms} filter length 
and a \unit[10]{ms} hop size. The architecture uses eight convolutional layers, one LSTM layer
and two fully-connected layers, each with ReLU activations except for the last layer, which has 
a sigmoid activation. The parameters used here are similar to the ones used 
in~\cite{wang2018voicefilter}, but now both the input and the output have two channels
with real and imaginary components, respectively. The prediction serves as a complex mask,  
consisting of a real and imaginary mask. The training stage uses a multi-resolution STFT 
loss function~\cite{defossez2020real}, which is the sum of all STFT loss functions using different 
STFT parameters.

\noindent \textbf{DARCN.\ } 
DARCN~\cite{li2020recursive} is a recently proposed monaural speech enhancement technique that 
uses multiple stages and that combines dynamic attention and recursive learning. Experiments are 
conducted with the open-source code\footnote{https://github.com/Andong-Li-speech/DARCN}
using a non-causal, 3-stage configuration.

\noindent \textbf{Proposed multi-stage SA-TCN Systems.\ }
The proposed multi-stage SA-TCN systems are characterized by the number of
stages $K$ and the hyper-parameters $(H,B,R,L)$. Each $K$-stage
SA-TCN system uses an STFT with a \unit[32]{ms} Hanning-window, a
\unit[32]{ms} filter length and a \unit[16]{ms} hop size. As such, $F =
257$. The multi-stage SA-TCN systems are trained using 80 epochs of 4-second
utterances from the LibriSpeech corpus and using 100 epochs of
variable-length utterances from the VCTK corpus. The proposed multi-stage
SA-TCN systems are trained using the Adam optimizer~\cite{kingma2015adam}
with an initial learning rate of 0.0002. All models use a mini-batch of 16
utterances. For each mini-batch of 16 utterances from the VCTK corpus, the
longest utterance is determined and the other utterances are zero-padded to
obtain equal-length utterances.

\subsection{ASR Setup.\ }
The automatic speech recognition (ASR) experiments use a time-delay neural 
network-hidden Markov model (TDNN-HMM) hybrid chain
model~\cite{peddinti2015time}. The TDNN models long-term temporal dependencies with 
training times that are comparable to standard feed-forward DNNs.
The data is represented at different time points by adding a
set of delays to the input, which allows the TDNN to have a finite dynamic
response to the time series input data. This acoustic model is trained using
the Kaldi toolkit~\cite{povey2011kaldi} with the standard 
recipe\footnote{https://github.com/kaldi-asr/kaldi/tree/master/egs/librispeech/s5}.
The ASR acoustic models were trained using 960 hours from the LibriSpeech training
set. The word error rate (WER) was measured using the LibriSpeech 
\qq{test-clean} set.

\subsection{Evaluation Metrics}
The speech enhancement systems are evaluated using the commonly used wide-band \textem{perceptual
evaluation of speech quality} (PESQ) score~\cite{rix2001perceptual,ITU-P.862,ITU-P.863},
the \textem{short-time objective intelligibility} (STOI) score~\cite{taal2011algorithm},
the scale-invariant signal-to-distortion ratio (SI-SDR)~\cite{le2019sdr},
and the {CSIG}, {CBAK} and {COVL} scores. The {CSIG} score is a signal distortion mean opinion 
score, the {CBAK} score measures background intrusiveness, and the {COVL} score measures the 
speech quality. The automatic speech recognition performance is measured
by determining the word error rate (WER).

\section{Experimental Performance Results}%
\label{sec:results}%
Extensive experiments have been performed to determine the performance of
the proposed multi-stage SA-TCN speech enhancement systems, This section
first details the findings of the ablation studies, and then presents 
the performance results for the multi-stage SA-TCN systems.

\subsection{Ablation Studies}
Ablation studies were performed to fine-tune the multi-stage SA-TCN system's 
hyper-parameters $(H,B,R,L)$, and to analyze the effectiveness of the self-attention and fusion blocks.

The performance of 5-stage SA-TCN systems is measured in terms of PESQ and STOI scores for 
several hyper-parameter configurations. The results are listed in Table~\ref{tab:diff_configs}.
We observe that it is more effective to increase the number of channels (hyper-parameters $B$
and $H$) in each TCN block than to increase the number of TCN blocks per stack ($L$). 
For instance, when $R=2$ and $H$ and $B$ are doubled, the PESQ score improves from 2.59 
to 2.65 and the STOI score improves from 92.36 to 93.02. At the same time, using $L=8$ 
instead of $L=5$ causes a slight degradation of the PESQ score. The performance can also 
be improved significantly by increasing the number of stacks $R$. We determined the model 
size for the larger TCN with $R=3$ stacks and $L=8$ TCN blocks per stack, which accounts
for about 1.68\,M parameters. Each SA block has about 0.2\,M parameters and each fusion 
block has about 1.7\,M parameters. If we only consider models with less than 10 million 
parameters, the model where $(H,B,R,L) = (256,128,3,8)$ performs best.
We should also note that there is a trade-off between the performance 
and the model size. 

\begin{table}[ht]
 \centering
 \caption{Performance for Several 5-stage SA-TCN Configurations}
 \def\b#1{\firstbest{#1}}
 \def\s#1{\secondbest{#1}}
 \def\t#1{\thirdbest{#1}}
 \label{tab:diff_configs}
 \setlength\tabcolsep{6pt}
 \setlength{\extrarowheight}{2pt}
\begin{tabular}{c|c|c|c|c|c|c|c}
 \hline
 $R$ & $L$ & $H$ & $B$ & $P$ & model size  &   PESQ  &   STOI   \\ 
 \hline
  2  &  5  & 128 &  64 &  3  &    2.38\,M  &   2.59  &   92.36  \\ 
  2  &  5  & 256 & 128 &  3  &    5.19\,M  &\t{2.65} &   93.02  \\ 
 \hline
  2  &  8  & 128 &  64 &  3  &    2.90\,M  &   2.53  &   92.32  \\ 
  2  &  8  & 256 & 128 &  3  &    7.21\,M  &   2.64  &\t{93.05} \\ 
 \hline
  3  &  5  & 128 &  64 &  3  &    2.81\,M  &   2.61  &   92.67  \\ 
  3  &  5  & 256 & 128 &  3  &    6.88\,M  &\s{2.71} &\b{93.40} \\ 
 \hline
  3  &  8  & 128 &  64 &  3  &    3.59\,M  &   2.60  &   92.20  \\ 
  3  &  8  & 256 & 128 &  3  &    9.91\,M  &\b{2.73} &\s{93.37} \\ 
 \hline
 \multicolumn{8}{c}{\scriptsize\text{The best score in a column is bold-faced, the second best}} \\[-3pt]
 \multicolumn{8}{c}{\scriptsize\text{is \s{navy blue} and the third best is \t{dark pink}.}}
 \end{tabular}
\end{table}

Next, we investigate the impact of the number of stages $K$ on the
performance of a multi-stage SA-TCN speech enhancement system. The motivation for
employing multi-stage learning is that the initial prediction is refined by
the next stage. The results in Table~\ref{tab:substage} show that the
performance improves step-wise after each stage. For instance, when
comparing the first and the fifth stage, it shows that the PESQ score
improves from 2.60 to 2.73, and the STOI score improves from
\unit[93.08]{\%} to \unit[93.37]{\%}. We also observe that the PESQ score's
rate of improvement gradually decreases from 0.5 to 0.1, which suggests that
adding further stages has diminishing returns in terms of performance and
that a 5-stage SA-TCN system is likely close to the upper bound on performance
for this multi-stage TCN-based approach.
\begin{table}[ht]
 \centering
 \caption{Per-Stage PESQ and STOI Scores for a 5-Stage SA-TCN System}
 \label{tab:substage}
 \setlength\tabcolsep{6pt}
 \setlength{\extrarowheight}{2pt}
 \begin{tabular}{l|c|c}
  \hline
    Stage        &  PESQ  &  STOI   \\ 
  \hline
    stage\,1     &  2.60  &  93.08  \\ 
    stage\,2     &  2.65  &  93.10  \\ 
    stage\,3     &  2.70  &  93.22  \\ 
    stage\,4     &  2.72  &  93.33  \\ 
    stage\,5     &  2.73  &  93.37  \\ 
  \hline
 \end{tabular}
\end{table}

\begin{figure}[!ht]
\begin{minipage}[b]{0.99\linewidth}
  \centerline{\includegraphics[width=0.99\textwidth]{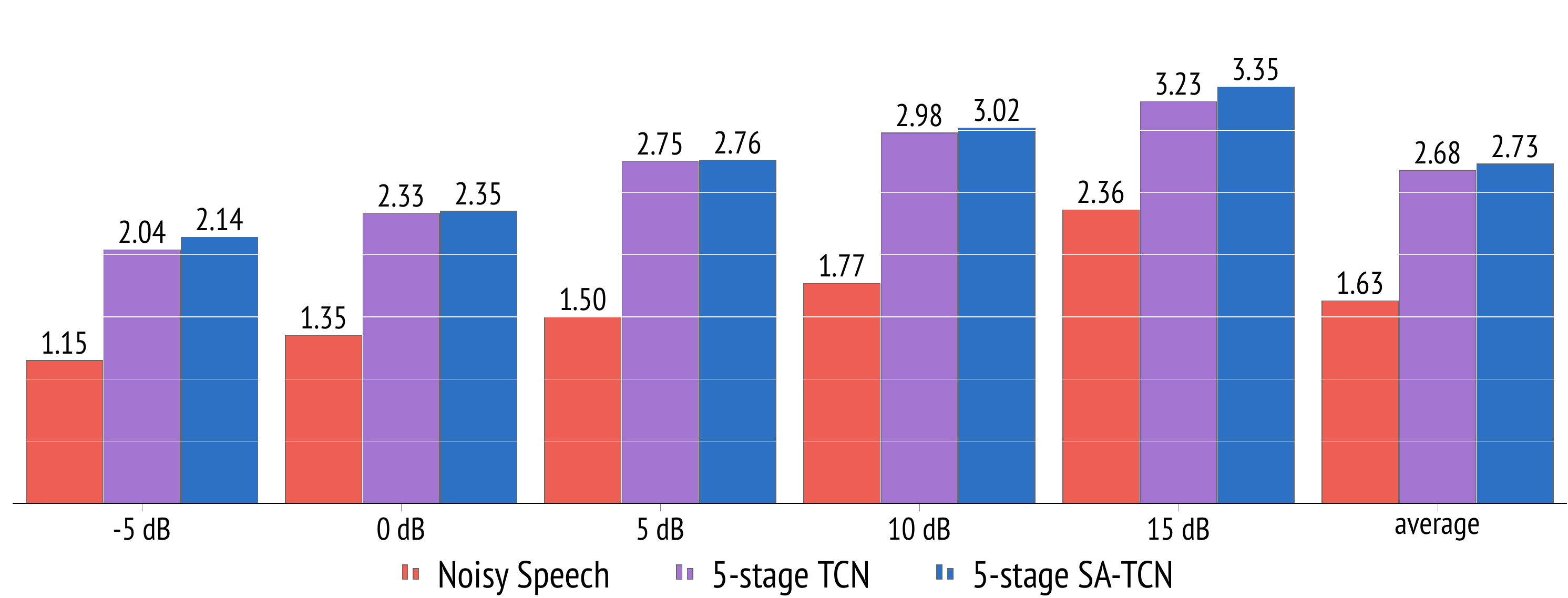}}
  \centerline{\footnotesize{\textem{a}.\  PESQ score}}\medskip
  \vspace{-2.5mm}
\end{minipage}
\hfill
\begin{minipage}[b]{0.99\linewidth}
  \centerline{\includegraphics[width=0.99\textwidth]{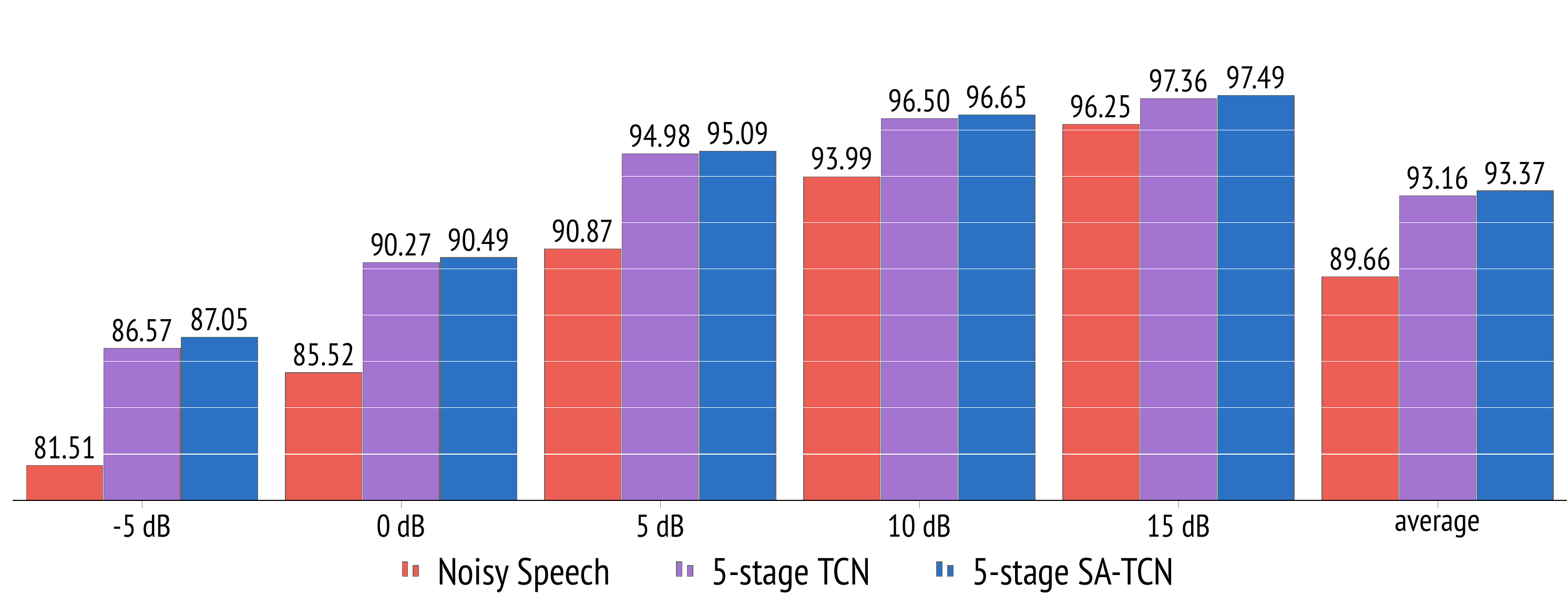}}
  \centerline{\footnotesize{\textem{b}.\ STOI score [\%]}}\medskip
\end{minipage}
\caption{%
  Impact of using a self-attention module in a 5-stage SA-TCN system, where 
  $(H,B,R,L) = (256,128,3,8)$.}
\label{fig:selfattn}
\end{figure}

The performance impact of using self-attention was determined using
PESQ and STOI scores. The results are shown in Fig.~\ref{fig:selfattn}. 
On average, a 5-stage SA-TCN system provides a STOI score improvement 
of \unit[3.5]{\%} and a PESQ score improvement of 1.05 relative to unprocessed 
noisy speech. The insertion of the SA block prior to the stacked layers of 
TCN blocks consistently improves PESQ and STOI scores for all SNR
conditions: the average PESQ score improves from 2.68 to 2.73 and the average 
STOI score improves from \unit[93.16]{\%} to \unit[93.37]{\%}. This indicates 
that the SA block is able to aggregate the frequency context, which is helpful 
for TCN-based speech enhancement. We also observe that the use of SA blocks 
show more significant performance gains at low SNR, e.g., at \unit[-5]{dB},
the PESQ score improves from 2.04 to 2.14 and the STOI score improves from \unit[86.57]{\%} 
to \unit[87.05]{\%}. This also indicates that multi-stage SA-TCN systems are 
more robust for lower SNR. 

\begin{figure}[!ht]
\begin{minipage}[b]{0.99\linewidth}
  \centerline{\includegraphics[width=0.99\textwidth]{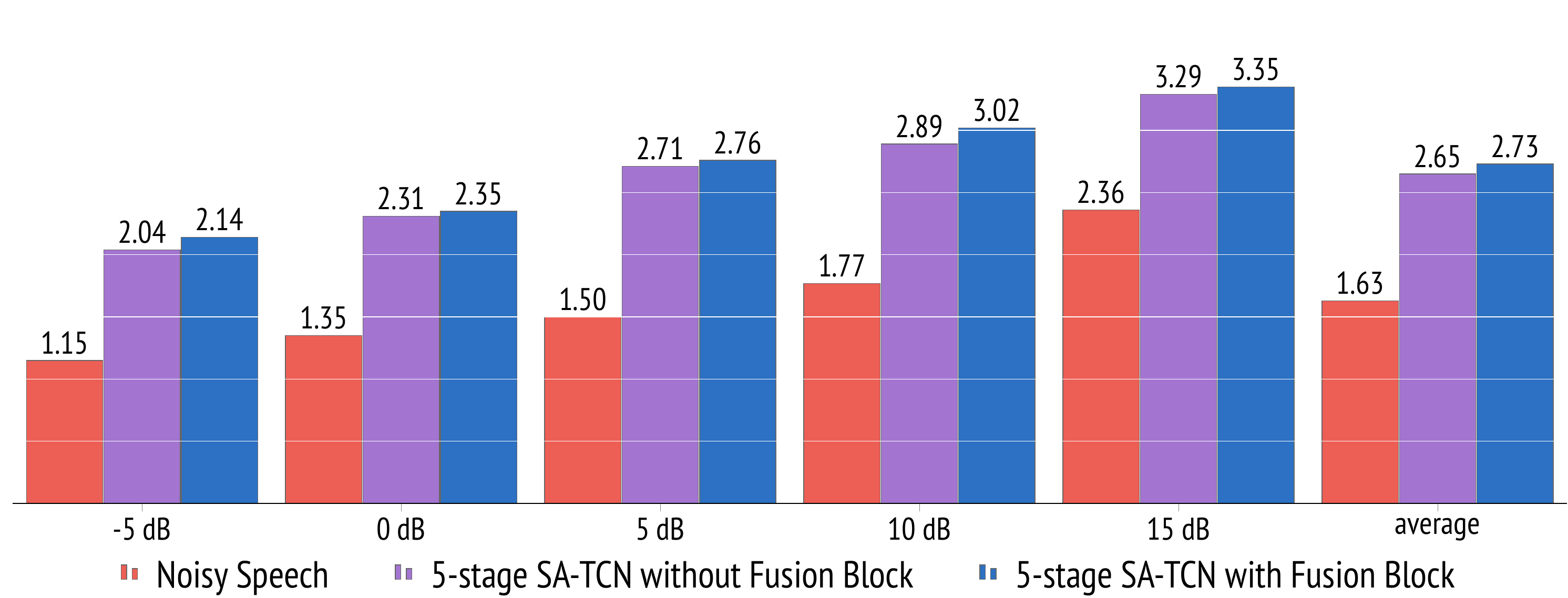}}
  \centerline{\footnotesize{\textem{a}.\  PESQ score}}\medskip
  \vspace{-2.5mm}
\end{minipage}
\hfill
\begin{minipage}[b]{0.99\linewidth}
  \centerline{\includegraphics[width=0.99\textwidth]{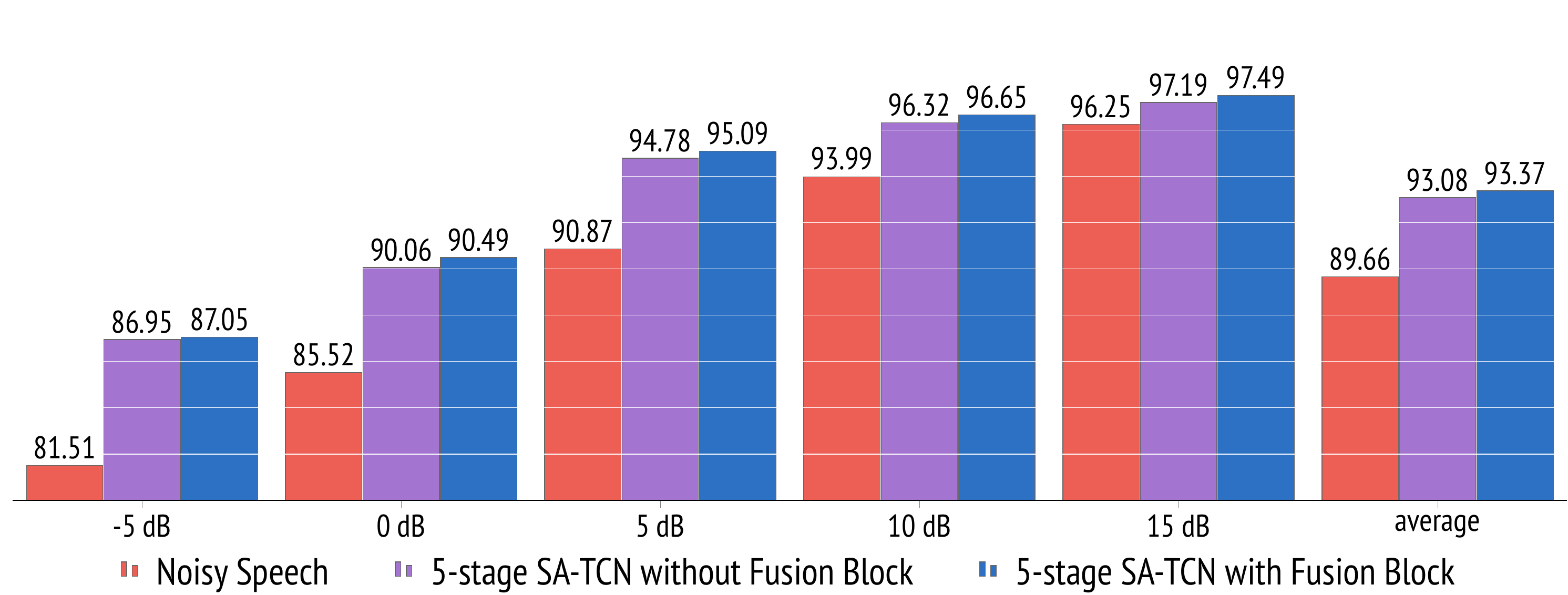}}
  \centerline{\footnotesize{\textem{b}.\ STOI score [\%]}}\medskip
\end{minipage}
\caption{%
  Impact of using a fusion block in a 5-stage SA-TCN system, where 
  $(H,B,R,L) = (256,128,3,8)$.}
\label{fig:fb}
\end{figure}

The effectiveness of the proposed fusion block, which re-injects original information 
in stages 3--5 in a 5-stage SA-TCN system to alleviate any speech signal
loss, is considered next. The PESQ and STOI scores are shown in Fig.~\ref{fig:fb}. 
It shows that both scores improve for all SNR scenarios. The average PESQ score
improves from 2.65 to 2.73, and the average STOI score improves from
\unit[93.08]{\%} to \unit[93.37]{\%}. The impact of the fusion block is, as
expected, more prominent at lower SNR, when the model not only removes the
noise, but can also easily partly remove the speech signal itself.

\subsection{Baseline System Comparison}
Extensive experiments with the proposed multi-stage SA-TCN system and the
CRN-based, complex-CNN and DARCN systems were conducted using the
LibriSpeech data set. All multi-stage SA-TCN systems use hyper-parameters $(H,B,R,L)
= (256, 128, 3, 8)$. Table~\ref{tab:pesq} shows that all multi-stage SA-TCN systems
outperform the baseline systems in terms of the PESQ score for the different
noise types and SNR conditions. The results also show that multi-stage SA-TCN systems with
more stages have a better PESQ score. Similarly, Table~\ref{tab:stoi} shows
that the STOI scores of the multi-stage SA-TCN systems are generally better than the
baseline systems, and that the best STOI scores are generally obtained for
4-stage and 5-stage SA-TCN systems. Interestingly, even the single-stage
SA-TCN system outperforms all baseline systems in terms of PESQ score.
Adding more stages improves the overall performance significantly. For
instance, the single-stage SA-TCN and the 2-stage SA-TCN have average PESQ 
scores of 2.47 and 2.67, respectively, and average STOI scores of 92.52\% and 92.88\%. The best
performance is achieved with $K=5$ stages, with an average PESQ score
of 2.73 and an average STOI score of \unit[93.37]{\%}. The proposed
5-stage SA-TCN system has much better PESQ and STOI scores than the baseline
systems, which demonstrates the effectiveness of the proposed approach. We
also observe that multi-stage learning is more effective at a low SNR. For
example, the 5-stage SA-TCN system achieves much better performance for Office
and Kitchen Noise at \unit[-5]{dB}, and it also performs well for Babble
Noise at low SNR.

Finally, we determined the SI-SDR metrics that quantify speech distortion.
Table~\ref{tab:si-sdr} shows that the proposed multi-stage SA-TCN
sytems generally outperform the baseline systems. We also observe that the SI-SDR
performance for multi-stage SA-TCN systems with $K>3$ stages decreases slightly,
which indicates that the additional stages not only mask the noise, but also
distort the speech signal. However, it will be shown next that these speech
distortions do not impact the ASR performance.
\begin{table*}[ht]
\centering
\def\b#1{\firstbest{#1}}
\def\s#1{\secondbest{#1}}
\def\t#1{\thirdbest{#1}}
\caption{PESQ Scores for Multi-Stage SA-TCN and Baseline Systems Using Samples from the LibriSpeech Corpus} 
\label{tab:pesq}
\setlength\tabcolsep{4pt}
\setlength{\extrarowheight}{2pt}
\begin{tabular}{l|ccccc|ccccc|ccccc|c}
\hline
 Noise type     &  \multicolumn{5}{c|}{Office Noise}              & \multicolumn{5}{c|}{Babble Noise}               &  \multicolumn{5}{c|}{Kitchen Noise}             & \multirow{2}{*}{Average} \\ \cline{1-16}
 SNR            &  -5     &   0     &   5     &  10     &  15     &  -5     &   0     &   5     &  10     &  15     &  -5     &   0     &   5     &  10     &  15     &         \\ 
\hline
 Noisy speech   &   1.30  &   1.68  &   2.01  &   2.60  &   3.39  &   1.06  &   1.09  &   1.22  &   1.37  &   1.80  &   1.07  &   1.18  &   1.30  &   1.62  &   2.10  &   1.63  \\
\hline
 CRN            &   1.90  &   2.16  &   2.58  &   3.07  &   3.45  &   1.08  &   1.20  &   1.46  &   1.75  &   2.24  &   1.43  &   1.81  &   2.19  &   2.44  &   2.78  &   2.09  \\
 Complex-CNN    &   2.14  &   2.40  &   2.84  &   3.01  &   3.24  &   1.13  &   1.30  &   1.70  &   2.06  &   2.54  &   1.77  &   2.20  &   2.46  &   2.67  &   2.95  &   2.30  \\
 DARCN          &   2.23  &   2.48  &   3.02  &   3.35  &   3.62  &   1.14  &   1.33  &   1.72  &   1.97  &   2.52  &   1.80  &   2.16  &   2.45  &   2.64  &   2.82  &   2.35  \\ 
\hline
 1-stage SA-TCN &   2.29  &   2.60  &   3.11  &   3.38  &   3.64  &   1.15  &   1.38  &   1.85  &   2.28  &   2.81  &   1.83  &   2.27  &   2.69  &   2.84  &   2.87  &   2.47  \\
 2-stage SA-TCN &   2.55  &   2.85  &\b{3.46} &\b{3.65} &\b{3.84} &\t{1.17} &\t{1.44} &   1.94  &   2.37  &   2.89  &   1.93  &\t{2.47} &\s{2.94} &   3.09  &\b{3.39} &   2.67 \\
 3-stage SA-TCN &\s{2.67} &\b{2.93} &\s{3.43} &\s{3.58} &\s{3.83} &\t{1.17} &   1.42  &\t{1.97} &\t{2.39} &\t{2.92} &\t{1.95} &\t{2.47} &\s{2.94} &\s{3.18} &\t{3.30} &\t{2.69}  \\
 4-stage SA-TCN &\t{2.61} &\s{2.90} &\t{3.42} &   3.56  &   3.81  &\b{1.19} &\s{1.46} &\s{2.03} &\s{2.48} &\s{2.95} &\s{1.97} &\s{2.49} &\b{2.98} &\t{3.11} &\s{3.35} &\s{2.70} \\
 5-stage SA-TCN &\b{2.74} &\s{2.90} &   3.38  &\t{3.57} &\t{3.82} &\s{1.18} &\b{1.47} &\b{2.05} &\b{2.51} &\b{3.02} &\b{2.10} &\b{2.53} &\s{2.94} &\b{3.19} &\t{3.30} &\b{2.73} \\ 
\hline
 \multicolumn{17}{c}{\scriptsize\text{All multi-stage SA-TCN models use hyper-parameters $(H,B,R,L) = (256, 128, 3, 8)$. The best score in a column is bold-faced, the second best}} \\[-3pt]
 \multicolumn{17}{c}{\scriptsize\text{is \s{navy blue} and the third best is \t{dark pink}.}}
\end{tabular}
\end{table*}

\begin{table*}[ht]
\centering
\def\b#1{\firstbest{#1}}
\def\s#1{\secondbest{#1}}
\def\t#1{\thirdbest{#1}}
\caption{STOI Scores for Multi-Stage SA-TCN and Baseline Systems Using Samples from the LibriSpeech Corpus} 
\label{tab:stoi}
\setlength\tabcolsep{4pt}
\setlength{\extrarowheight}{2pt}
\begin{tabular}{l|ccccc|ccccc|ccccc|c}
\hline
 Noise type     &   \multicolumn{5}{c|}{Office}                   &   \multicolumn{5}{c|}{Babble}                   &   \multicolumn{5}{c|}{Kitchen}            &   \multirow{2}{*}{Average} \\ \cline{1-16}
 SNR [dB]       &   -5    &    0    &   5     &   10    &   15    &   -5    &    0    &    5    &   10    &   15    &   -5    &    0    &    5    &   10    &   15    &          \\ 
\hline
 Noisy speech   &   92.91 &   96.81 &   98.50 &   98.62 &   99.25 &   54.94 &   64.93 &   80.04 &   87.10 &   91.08 &   84.59 &   91.75 &   95.40 &   98.21 &   98.86 &   89.66  \\ 
\hline
 CRN            &   93.02 &   96.15 &   97.29 &   97.55 &   98.07 &   54.44 &   68.15 &   83.75 &   90.04 &   91.98 &   87.14 &   92.60 &   95.57 &   97.15 &   97.23 &   90.23  \\
 Complex-CNN    &   93.23 &   95.78 &   97.25 &   96.91 &   97.01 &   60.70 &   72.92 &   87.01 &   91.85 &   93.26 &   87.38 &   92.63 &   95.11 &   96.83 &   96.83 &   91.09  \\
 DARCN          &\s{95.08}&   97.04 &   98.46 &\t{98.44}&   98.82 &   62.60 &   75.20 &   88.90 &   91.72 &   94.41 &\s{90.31}&   93.71 &   96.60 &\b{98.31}&   98.31 &   92.64  \\ 
\hline
 1-stage SA-TCN &   94.81 &   97.09 &\s{98.55}&\s{98.52}&   98.76 &   61.51 &   74.96 &   88.62 &   92.89 &   94.26 &   89.37 &   94.11 &   96.37 &   98.08 &   98.27 &   92.52  \\
 2-stage SA-TCN &   94.85 &   96.98 &   98.35 &   98.24 &\t{98.89}&\t{64.91}&\t{76.59}&   89.89 &   93.29 &   94.47 &\t{89.45}&\t{93.72}&\t{96.76}&   97.67 &\s{98.65}&\t{92.88} \\
 3-stage SA-TCN &\t{95.02}&\s{97.23}&   98.48 &   98.32 &\s{98.92}&   63.30 &   75.75 &\t{89.92}&\t{93.63}&\t{94.64}&   88.71 &   93.68 &\s{96.79}&\t{98.12}&   98.60 &   92.82  \\
 4-stage SA-TCN &   94.93 &\t{97.14}&\t{98.52}&   98.22 &   98.75 &\b{65.47}&\s{76.94}&\b{90.33}&\s{93.70}&\s{94.73}&   89.41 &\s{94.47}&   96.74 &   98.10 &\t{98.62}&\s{93.10} \\
 5-stage SA-TCN &\b{95.43}&\b{97.24}&\b{98.62}&\b{98.60}&\b{98.96}&\s{64.26}&\b{77.57}&\b{90.33}&\b{93.94}&\b{94.94}&\b{90.82}&\b{94.84}&\b{96.88}&\s{98.22}&\b{98.78}&\b{93.37} \\ 
\hline
 \multicolumn{17}{c}{\scriptsize\text{All multi-stage SA-TCN models use hyper-parameters $(H,B,R,L) = (256, 128, 3, 8)$. The best score in a column is bold-faced, the second best}} \\[-3pt]
 \multicolumn{17}{c}{\scriptsize\text{is \s{navy blue} and the third best is \t{dark pink}.}}
\end{tabular}
\end{table*}

\begin{table}[ht]
\def\b#1{\firstbest{#1}}
\def\s#1{\secondbest{#1}}
\def\t#1{\thirdbest{#1}}
\caption{SI-SDR scores for Multi-Stage SA-TCN and Baseline Systems Using Samples from the LibriSpeech Corpus}
\label{tab:si-sdr}
\setlength\tabcolsep{4pt}
\setlength{\extrarowheight}{2pt}
\begin{center}
\begin{tabular}{l|c|c|c|c|c|c}
\hline
 SNR [dB]        &   -5    &    0    &    5    &   10    &   15    &  Average \\ 
\hline
 CRN             &    1.43 &    5.72 &    9.63 &   13.07 &   16.39 &    9.28  \\ 
 Complex-CNN     &    9.66 &   11.32 &   13.35 &   14.49 &   16.86 &   13.15  \\
 DARCN           &   11.00 &   12.65 &   15.84 &\t{17.60}&\s{19.87}&   15.41  \\ 
\hline
 1-stage SA-TCN  &\t{11.13}&\b{13.40}&\s{16.77}&\s{18.05}&\t{19.77}&\s{15.84} \\ 
 2-stage SA-TCN  &   10.80 &\t{12.71}&\t{16.60}&   17.42 &   19.44 &   15.42  \\ 
 3-stage SA-TCN  &\s{11.41}&\s{13.03}&\b{17.01}&\b{18.34}&\b{20.33}&\b{16.04} \\
 4-stage SA-TCN  &   10.48 &   12.69 &   16.25 &   17.10 &   19.51 &   15.23  \\
 5-stage SA-TCN  &\b{11.43}&   12.82 &   16.29 &   17.41 &   19.67 &\t{15.55} \\ 
\hline
\multicolumn{7}{c}{\scriptsize\text{The best score in a column is bold-faced, the second best}} \\[-3pt]
\multicolumn{7}{c}{\scriptsize\text{is \s{navy blue} and the third best is \t{dark pink}.}}
\end{tabular}
\end{center}
\end{table}

\subsection{Automatic Speech Recognition}
We conducted automatic speech recognition (ASR) experiments using LibriSpeech to assess 
the performance of multi-stage SA-TCN systems with up to five stages and determined the 
word error rate (WER) as well as the WER reduction. The baseline systems are the
CRN-based method, and the complex-CNN and DARCN methods. The results are shown in 
Table~\ref{tab:asr}. Our 1-stage SA-TCN system performs slightly worse than the 
best baseline systems, but the multi-stage SA-TCN methods perform better, and the 
the proposed 5-stage SA-TCN achieves an absolute improvement of
\unit[18.8]{\%}, \unit[8.4]{\%} and \unit[4.6]{\%} 
relative to CRN, complex-CNN and the DARCN methods, respectively. The ASR results are 
similar to the STOI performance.
\begin{table}[ht]
 \def\b#1{\firstbest{#1}}
 \def\s#1{\secondbest{#1}}
 \def\t#1{\thirdbest{#1}}
\centering
 \caption{Speech Recognition Performance of Multi-Stage SA-TCN and Baseline Systems}
  \label{tab:asr}
  \setlength\tabcolsep{4pt}
  \setlength{\extrarowheight}{2pt}
  \begin{tabular}{l|c|c}
   \hline
    Method         & WER [\%] & WER reduction [\%] \\
   \hline
    Noisy Speech   &   32.94  &     --    \\ 
    CRN            &   30.86  &    6.3    \\ 
    Complex-CNN    &   27.44  &   16.7    \\ 
    DARCN          &   26.18  &   20.5    \\ 
   \hline
    1-stage SA-TCN &   27.86  &   15.4    \\ 
    2-stage SA-TCN &\t{25.32} &\t{23.1}   \\ 
    3-stage SA-TCN &   26.11  &   20.7    \\ 
    4-stage SA-TCN &\s{25.27} &\s{23.3}   \\ 
    5-stage SA-TCN &\b{24.67} &\b{25.1}   \\ 
   \hline
   \multicolumn{3}{c}{\scriptsize\text{The best score in a column is bold-faced, the second best}} \\[-3pt]
    \multicolumn{3}{c}{\scriptsize\text{is \s{navy blue} and the third best is \t{dark pink}.}}
  \end{tabular}
\end{table}

\subsection{Spectrogram-Based Visualization}
\begin{figure*}[!ht]
\begin{minipage}[b]{0.3\linewidth}
  \centering
  \centerline{\includegraphics[height=3.8cm]{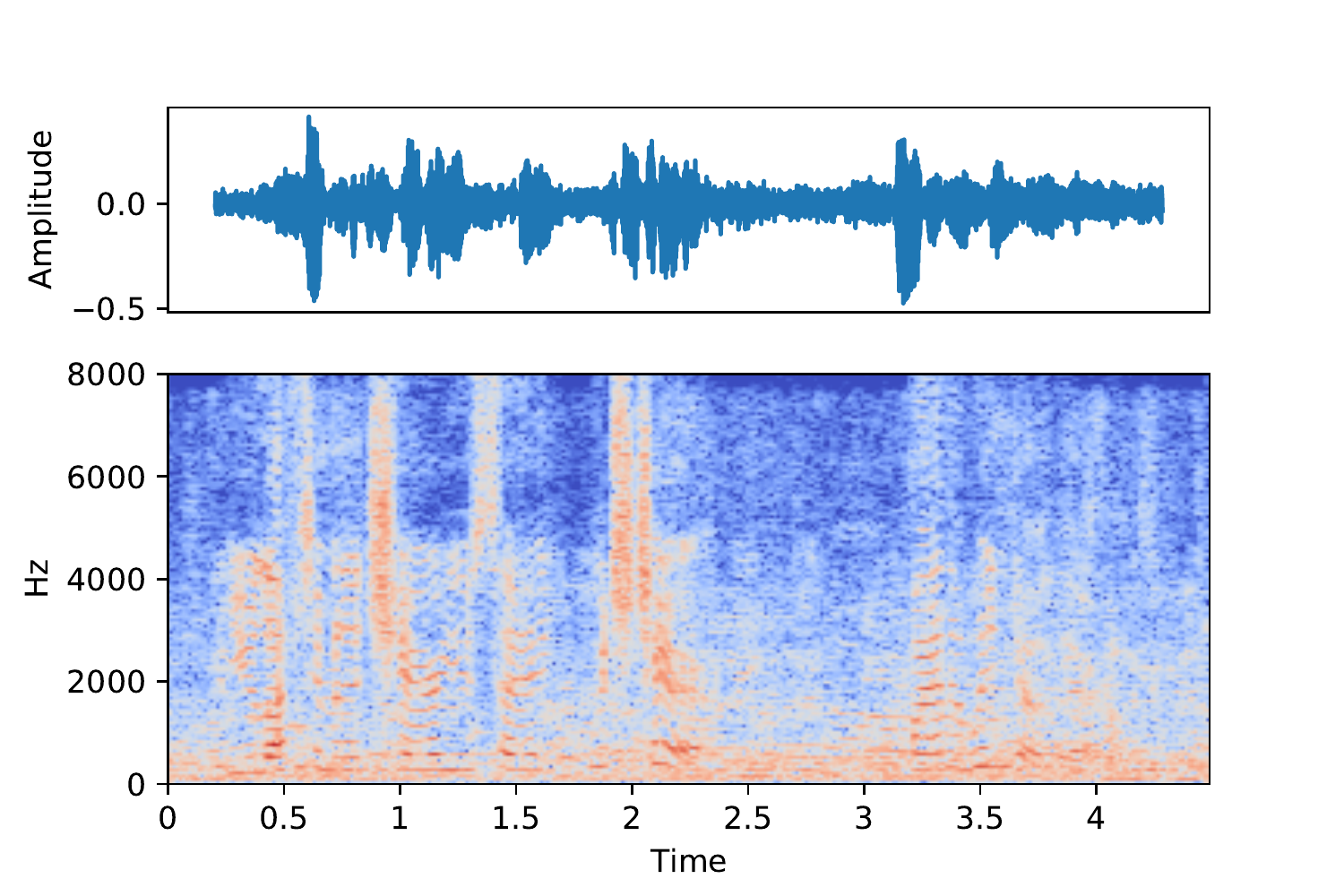}}
  \vspace{-1.5ex}%
  \centerline{\scriptsize{\textem{a.}\ Noisy Speech}}\smallskip
\end{minipage}
\hfill
\begin{minipage}[b]{0.3\linewidth}
  \centering
  \centerline{\includegraphics[height=3.8cm]{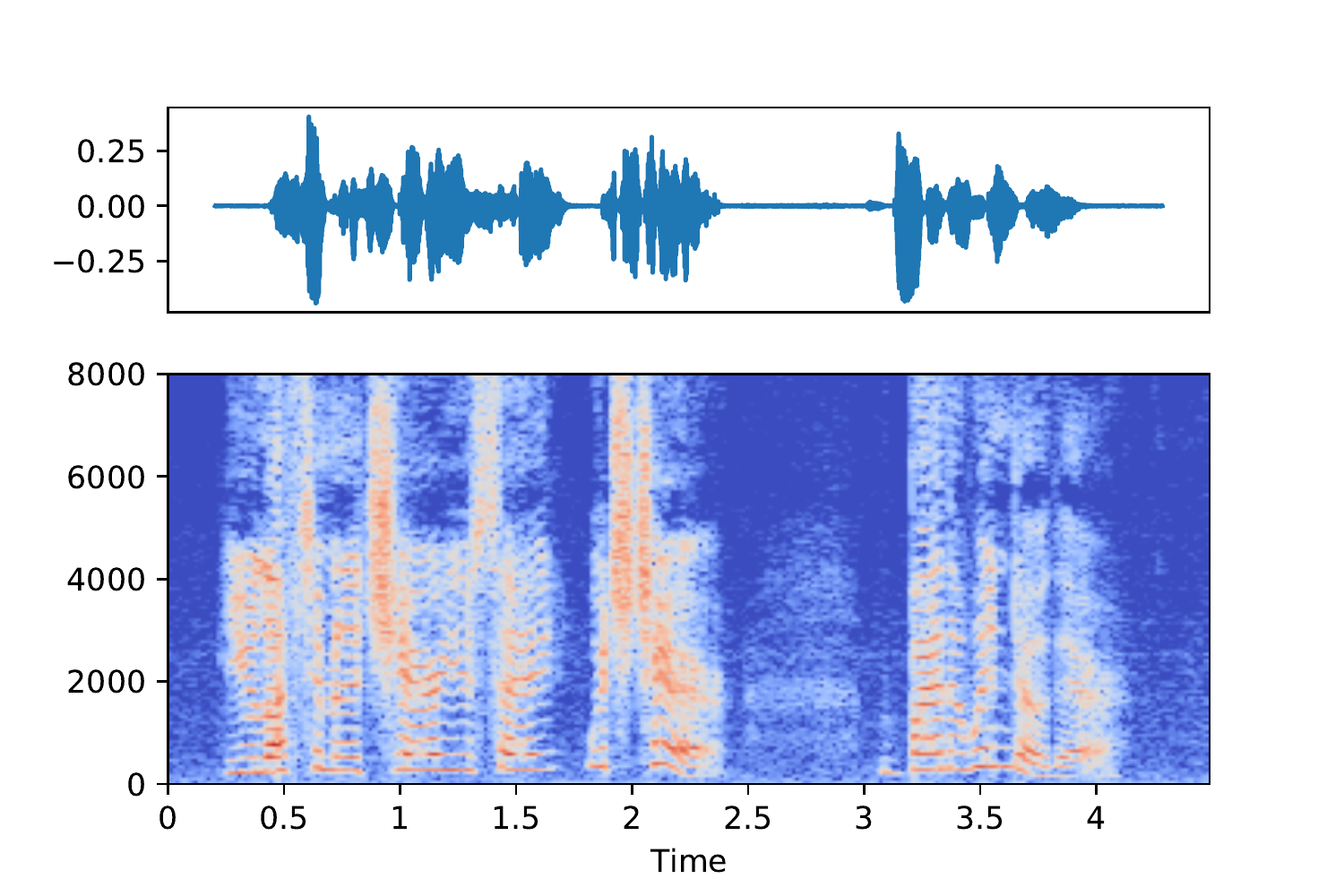}}
  \vspace{-1.5ex}%
  \centerline{\scriptsize{\textem{b.}\ Clean Target}}\smallskip
\end{minipage}
\hfill
\begin{minipage}[b]{0.3\linewidth}
  \centering
  \centerline{\includegraphics[height=3.8cm]{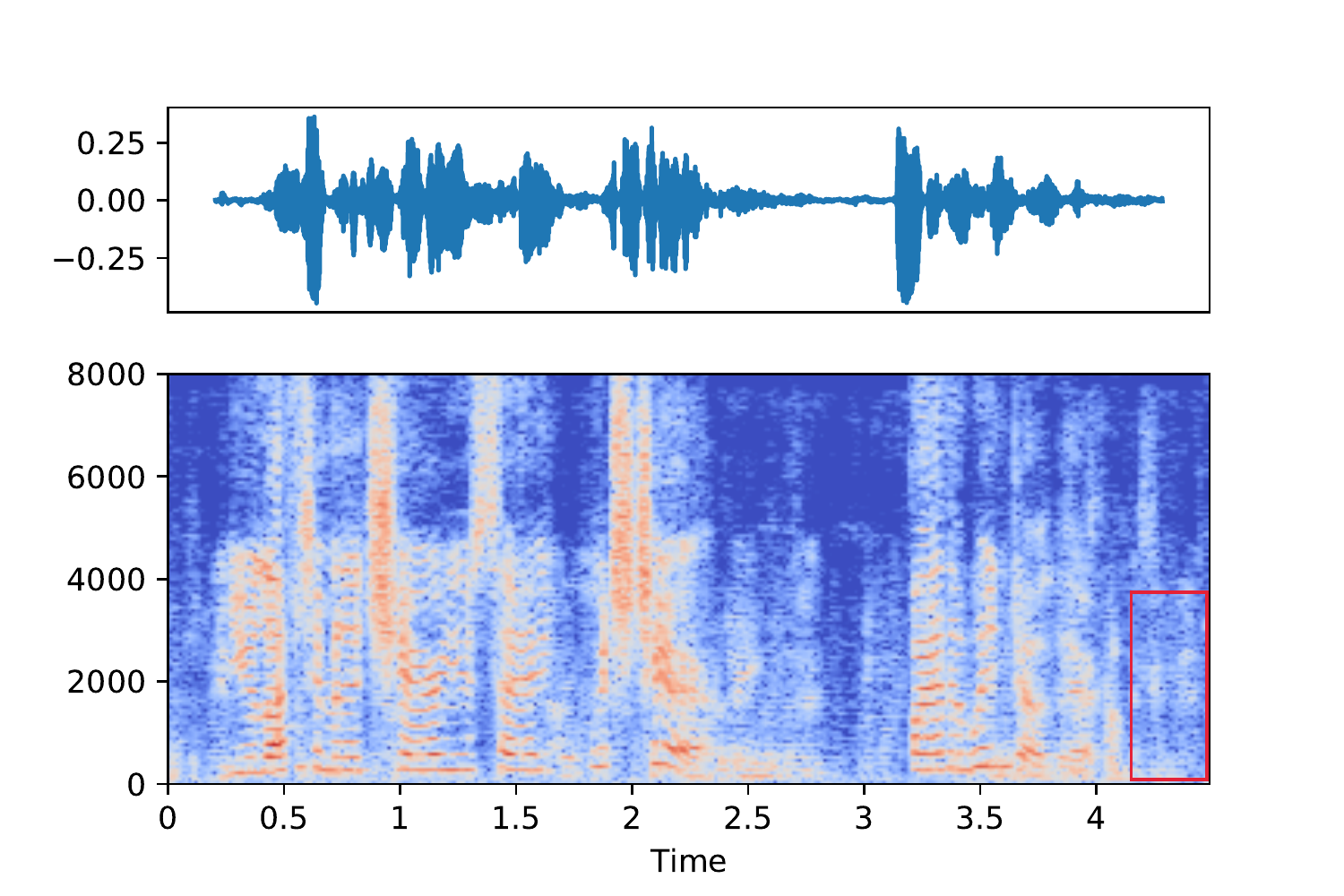}}
  \vspace{-1.5ex}%
  \centerline{\scriptsize{\textem{c.}\ CRN-based System}}\smallskip
\end{minipage}
\hfill
\begin{minipage}[b]{0.3\linewidth}
  \centering
  \centerline{\includegraphics[height=3.8cm]{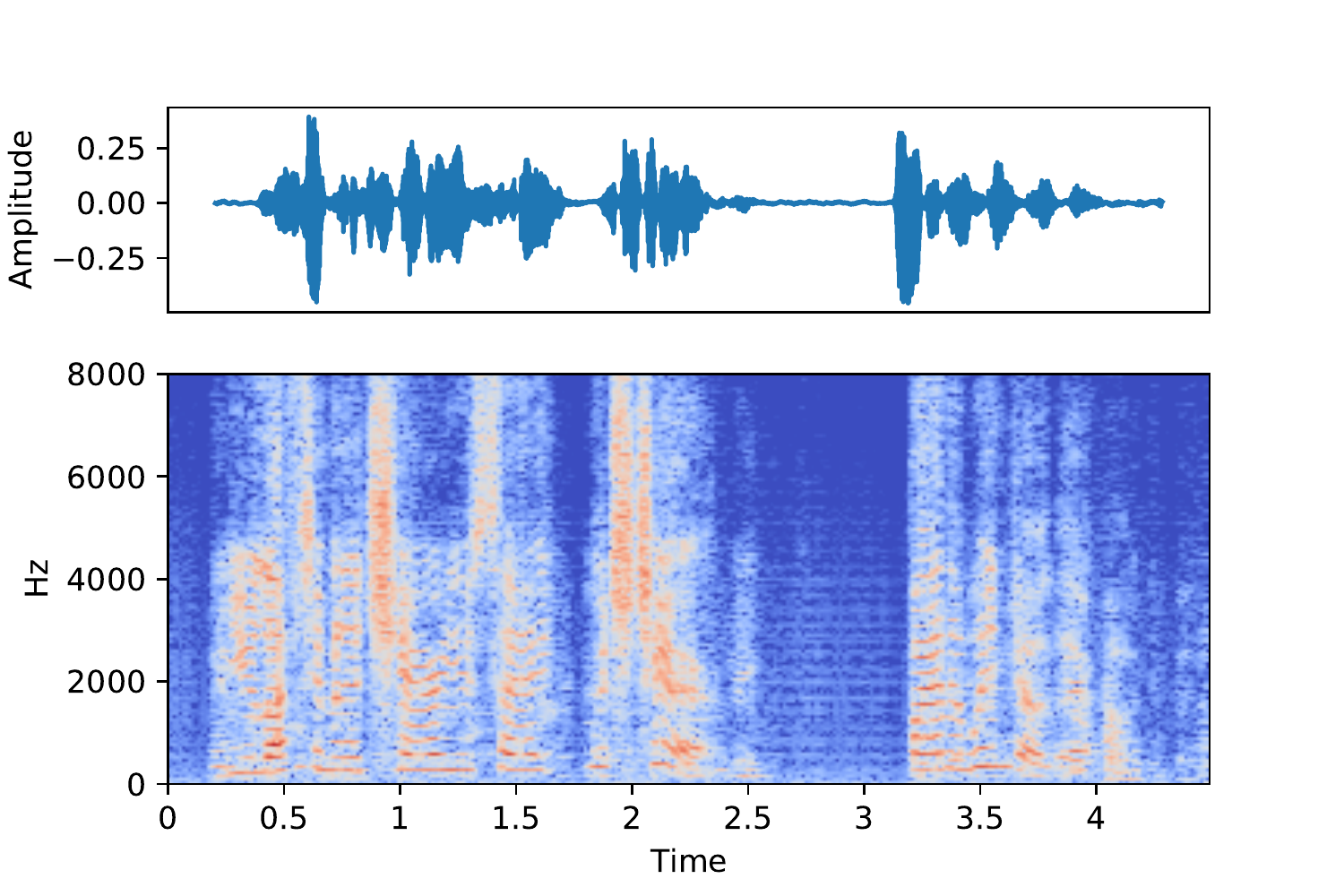}}
  \vspace{-1.5ex}
  \centerline{\scriptsize{\textem{d.}\ Complex-CNN System}}\smallskip
\end{minipage}
\hfill
\begin{minipage}[b]{0.3\linewidth}
  \centering
  \centerline{\includegraphics[height=3.8cm]{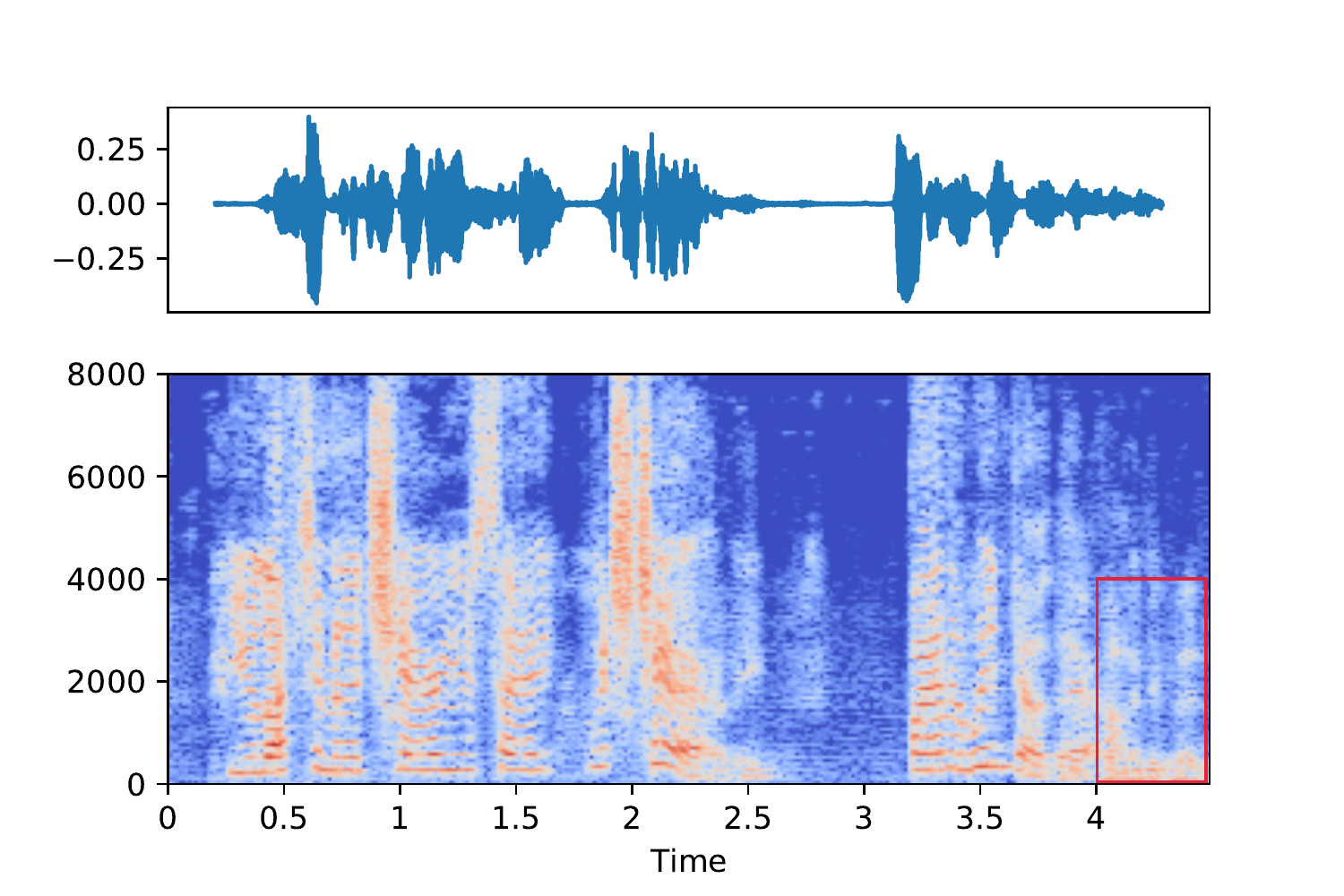}}
  \vspace{-1.5ex}%
  \centerline{\scriptsize{\textem{e.}\ DARCN System}}\smallskip
\end{minipage}
\hfill
\begin{minipage}[b]{0.3\linewidth}
  \centering
  \centerline{\includegraphics[height=3.8cm]{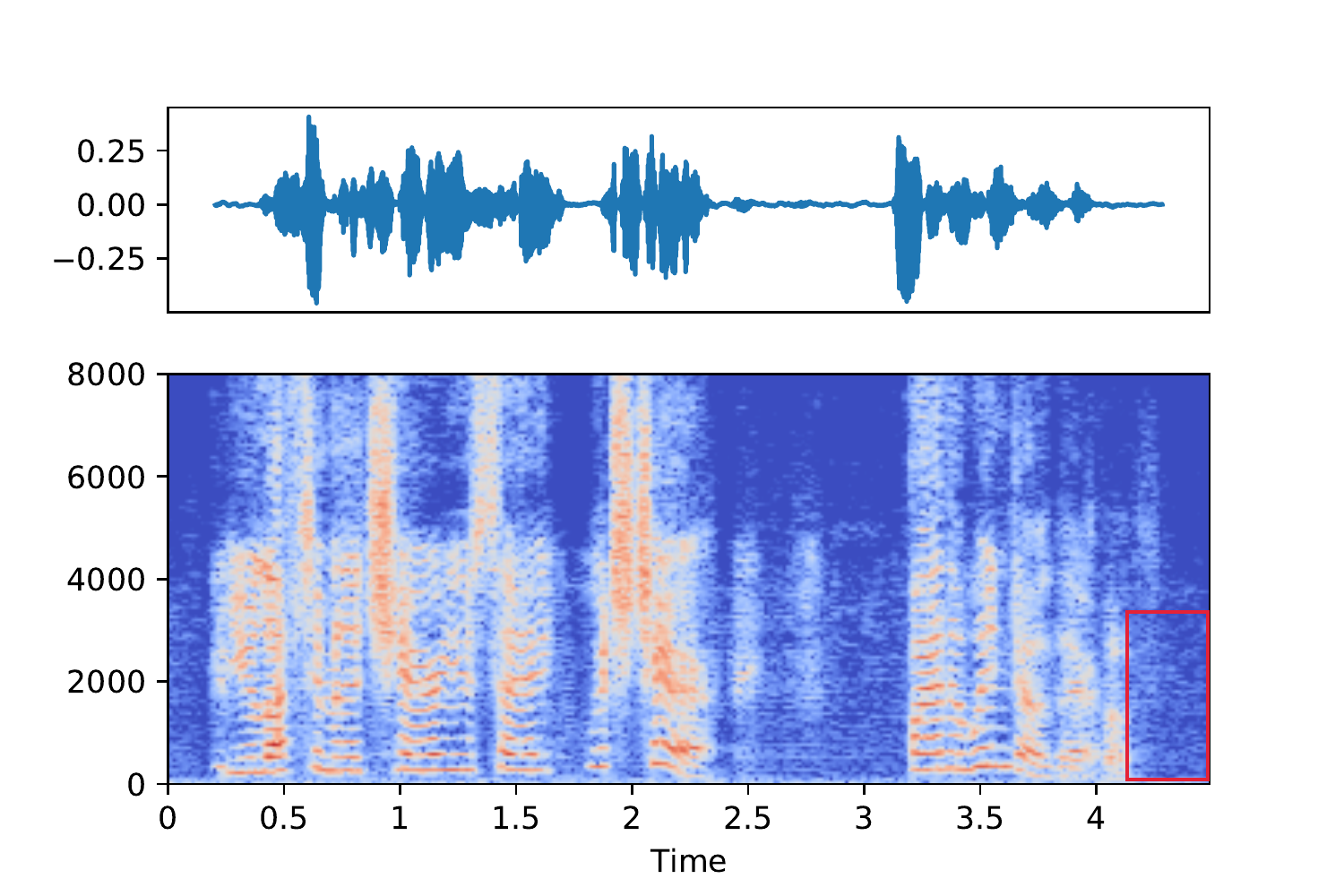}}
  \vspace{-1.5ex}%
  \centerline{\scriptsize{\textem{f.}\ 5-Stage SA-TCN System}}\smallskip
\end{minipage}
\caption{Spectrograms of a sample input mixed with Babble 
    Noise at an SNR of \unit[5]{dB} and the speech-enhanced 
    signals at the output of SA-TCN and baseline systems.}
\label{fig:visualization}
\end{figure*}

Speech enhancement performance can be assessed using spectrograms.
Consider the situation where clean speech is perturbed by Babble 
noise at an SNR of \unit[5]{dB}. Fig.~\ref{fig:visualization} shows
spectrograms of the noisy speech signal, the clean speech 
target, as well as the CRN-based and complex CNN-based systems, 
the DARCN system, and the proposed 5-stage SA-TCN enhanced speech system.
The spectrograms clearly show that the proposed system 
is best at suppressing residual noise while preserving the speech patterns.

\subsection{Speech-Enhancement Benchmark Results}
The proposed multi-stage SA-TCN speech enhancement systems are compared with
state-of-the-art methods using the publicly available benchmark data set
VCTK. As shown in Table~\ref{tab:vctk}, the proposed multi-stage SA-TCN
systems outperform methods that use T-F frequency features, including magnitude,
gamma-tone spectral and complex STFT in terms of all the speech enhancement
metrics used in this paper. Compared with the recently proposed time-domain
method DEMUCS, our proposed method uses fewer parameters and achieves better
performance in terms of CBAK and COVL metrics, while the PESQ, STOI and CSIG
are slightly worse. The experiments with the VCTK corpus show that adding
more stages still provides some incremental performance improvements.
\begin{table*}[!ht]
 \centering
  \def\b#1{\firstbest{#1}}
 \def\s#1{\secondbest{#1}}
 \def\t#1{\thirdbest{#1}}
 \caption{Performance Evaluation Scores of Multi-Stage SA-TCN and Baseline Systems Using Samples 
          from the VCTK Corpus} 
 \label{tab:vctk}
 \setlength\tabcolsep{4pt}
 \setlength{\extrarowheight}{2pt}
 \begin{tabular}{l|c|c|cccccc}
   \hline
                                                     & model size & feature type        &   PESQ &   STOI   &   CSIG  &   CBAK  &   COVL  &   SI-SDR  \\ 
   \hline
     noisy speech                                    &      --    & --                  &   1.97 &   0.921  &   3.35  &   2.44  &   2.63  &    8.45   \\ 
   \hline
     SEGAN~\cite{pascual2017segan} (2017)            &   43.2\,M  & Waveform            &   2.16 &   0.93   &   3.48  &   2.94  &   2.80  &      --   \\
     Wave-U-Net~\cite{macartney2018improved} (2018)  &   10.2\,M  & Waveform            &   2.40 &     --   &   3.52  &   3.24  &   2.96  &      --   \\
     DFL~\cite{germain2018speech} (2018)             &   0.64\,M  & Waveform            &     -- &     --   &   3.86  &   3.33  &   3.22  &      --   \\
     MMSE-GAN~\cite{soni2018time} (2018)             &   0.79\,M  & Gamma-tone spectral &   2.53 &   0.93   &   3.80  &   3.12  &   3.14  &      --   \\
     MetricGAN~\cite{fu2019metricgan} (2019)         &   1.89\,M  & Magnitude           &   2.86 &          &   3.99  &   3.18  &   3.42  &      --   \\
     MB-TCN~\cite{zhang2019monaural} (2019)          &   1.66\,M  & Magnitude           &   2.94 &   0.9364 &   4.21  &   3.41  &   3.59  &      --   \\
     DeepMMSE~\cite{zhang2020deepmmse} (2020)        &     --     & Magnitude           &   2.95 &   0.94   &\t{4.28} &   3.46  &   3.64  &      --   \\
     MHSA-SPK~\cite{koizumi2020speech} (2020)        &     --     & STFT                &   2.99 &     --   &   4.15  &   3.42  &   3.57  &      --   \\
     STFT-TCN~\cite{koyama2020exploring} (2020)      &     --     & STFT                &   2.89 &     --   &   4.24  &   3.40  &   3.56  &      --   \\
     DEMUCS~\cite{defossez2020real} (2020)           &  127.9\,M  & Waveform            &\b{3.07}&\b{0.95}  &\b{4.31} &   3.40  &   3.63  &      --   \\ 
   \hline
     1-stage SA-TCN                                  &   1.88\,M  & Magnitude           &   2.84 &   0.9402 &   4.16  &   3.37  &   3.50  &   17.98   \\
     2-stage SA-TCN                                  &   3.76\,M  & Magnitude           &   2.96 &   0.9422 &   4.25  &   3.45  &   3.62  &   18.19   \\
     3-stage SA-TCN                                  &   5.81\,M  & Magnitude           &   2.99 &   0.9423 &   4.27  &\t{3.48} &\t{3.64} &\t{18.38}  \\  
     4-stage SA-TCN                                  &   7.86\,M  & Magnitude           &\t{3.01}&\t{0.9428}&   4.27  &\s{3.49} &\s{3.66} &\s{18.40}  \\
     5-stage SA-TCN                                  &   9.91\,M  & Magnitude           &\s{3.02}&\s{0.9439}&\s{4.29} &\b{3.50} &\b{3.67} &\b{18.48}  \\     
   \hline
    \multicolumn{9}{c}{\scriptsize\text{The best score in a column is bold-faced, the second best is \s{navy blue} and the third best is \t{dark pink}.}}
  \end{tabular}
\end{table*}

\section{Discussion and Conclusions}%
\label{sec:concl}%
In this paper, we have presented novel multi-stage SA-TCN speech enhancement
systems, where each stage consists of a self-attention block followed by $R$
stacks of $L$ temporal convolutional network blocks with doubling dilation
factors. The stacks of $L$ TCN blocks effectively perform sequential
refinement processing. Multi-stage SA-TCN systems with three or more stages
use a fusion block as of the third stage to mitigate any possible loss of
the original speech information loss in later stages. The proposed
self-attention module is used to provide a dynamic representation by
aggregating the frequency context. Extensive experiments were used to
fine-tune the hyper-parameters. It was shown that both the addition of 
the self-attention modules and the fusion blocks resulted in better 
performance. We noted that even the basic 1-stage SA-TCN system performs 
well and that adding stages improves the speech enhancement 
scores. The model size increases almost linearly with the number of 
stages. The relative improvement when adding an additional stage reduces
when more stages are added and as such one approaches an implicit upper
bound for this approach. The best overall performance with a reasonable 
model size was obtained with a 5-stage SA-TCN system.

Extensive experiments were conducted using the LibriSpeech and VCTK data
sets to determine the performance of the multi-stage SA-TCN speech
enhancement systems and to compare the proposed system with other
state-of-the-art deep-learning speech enhancement systems. It was shown that
the proposed multi-stage SA-TCN methods achieve better performance in terms
of widely used objective metrics while having fewer parameters. Speech
enhancement, especially in mobile applications, requires computational- and
parameter-efficient models. The proposed methods meet this requirement and
at the same time provide excellent performance. Spectrograms were used to
visualize that the proposed 5-stage SA-TCN systems can remove noise
effectively while preserving the speech patterns. The proposed multi-stage
SA-TCN systems predict a soft mask at each stage, which can be viewed as an
implicit ideal ratio mask (IRM).
For speech signals that are dominated by noise, the noise is suppressed 
gradually in each stage, which is a main reason for the excellent 
performance. The proposed multi-stage SA-TCN systems are also shown to have 
excellent ASR performance.

The focus of this paper is to process and enhance the spectrum magnitude,
and the unaltered noisy phase is used when reconstructing the waveforms in
the time domain. Recently, several studies have shown that phase information
is also important for improving the perceptual quality~\cite{paliwal2011importance, 
li2020two}. Thus, incorporating phase information into the proposed approach
may lead to further improvements. This is currently being investigated. 

\section*{Acknowledgments}
The authors thank Clemson University for the generous allotment of compute time on
its Palmetto cluster.

\bibliographystyle{IEEEtran}
\bibliography{IEEEabrv,MS_SA_TCN}

\end{document}